\documentclass[12pt]{article}
\pdfoutput=1
\usepackage{array}
\usepackage{graphicx}
\usepackage{amssymb}
\usepackage{amsmath}
\usepackage{latexsym}
\usepackage[usenames]{color}
\usepackage{cite}
\usepackage{wasysym}
\definecolor{darkblue}{cmyk}{0.9,0.9,0,0}
\definecolor{darkred}{rgb}{0.6,0,0.3}
\usepackage{graphicx} 

\usepackage[setpagesize=false,pagebackref=false, linktocpage, colorlinks=true, linkcolor=darkblue,citecolor=darkblue,urlcolor=darkblue]{hyperref}

\renewcommand{\thefootnote}{\arabic{footnote}}
\allowdisplaybreaks

\def\eqref#1{(\ref{#1})}

\newcommand{\beq}{\begin{equation}}
\newcommand{\eeq}{\end{equation}}

\begin{document}
\thispagestyle{empty}

\renewcommand{\thefootnote}{\fnsymbol{footnote}}
\setcounter{page}{1}
\setcounter{footnote}{0}
\setcounter{figure}{0}
\begin{center}
$$$$
{\Large\textbf{\mathversion{bold}
Decagon at Two Loops}\par}

\vspace{1.3cm}

\textrm{Thiago Fleury$^{\textcolor[rgb]{0,0.6,0}{\triangledown}}$, Vasco Goncalves$^{\textcolor[rgb]{0.9,0,0}{\hexagon}}$}
\\ \vspace{1.2cm}
\footnotesize{\textit{
$^{\textcolor[rgb]{0,0.6,0}{\triangledown}}$ 
International Institute of Physics, Federal University of Rio Grande do Norte, \\ Campus Universit\'ario,  Lagoa Nova, Natal, RN 59078-970, Brazil
\vspace{3mm} \\
$^{\textcolor[rgb]{0.8,0,0}{\hexagon}}$ 
Instituto de F\'isica Te\'orica, UNESP - Univ. Estadual Paulista, 
ICTP South American Institute for Fundamental Research,
Rua Dr. Bento Teobaldo Ferraz 271, 01140-070, S\~ao Paulo, SP, Brasil
}  
\vspace{4mm}
}

\par\vspace{1.5cm}

\textbf{Abstract}\vspace{2mm}
\end{center}

We have computed the simplest five point function in $\mathcal{N}=4$ SYM at two loops using the hexagonalization approach to correlation functions. Along the way we have determined all two-particle mirror contributions at two loops and we have computed all the integrals involved in the final result. As a test of our results we computed a few four-point functions
and they agree with the perturbative results computed previously.  We have also obtained $l$ loop results for some parts of the two-particle contributions
with $l$ arbitrary. We also derive differential equations for a class of integrals that should appear at higher loops in the five point function.


\noindent

\setcounter{page}{1}
\renewcommand{\thefootnote}{\arabic{footnote}}
\setcounter{footnote}{0}
\setcounter{tocdepth}{2}
\newpage
\tableofcontents

\parskip 5pt plus 1pt   \jot = 1.5ex

\section{Introduction\label{sec:intro}}
Correlation functions of local operators are one of the most interesting physical observables that can be considered in a conformal field theory. Corrrelators 
of half-BPS operators in $\mathcal{N}=4$ SYM have seen in the last two decades an abundance of new results for weak \cite{Edens1,weakI,weakII,weakIII,DrukkerPlefka,EdensII,noteweak,3loopdata,45loopdata}, strong \cite{strongI,strongII,strongIII,strongIV,strongV,strongVI,strongVII,strongVIII,strongIX,strongX,strongXI,strongXII}  and in some cases finite coupling regime
\cite{Frank1,Frank2,KostovOctagon,KostovOctagon2,OctagonsI,OctagonsII,NullOctagon,NullOctagonII}. However, the results for more than four operators are scarce and in most cases limited to one loop computation which due to the nature of interactions of $\mathcal{N}=4$ SYM does not explore the full complexity of the problem. 
Two exceptions are the two-loop computations of five and six point correlators 
of length two half-BPS operators of 
\cite{Edenfive,Edensix} and 
the recent supergravity computation of a five point function\cite{Goncalves:2019znr}. 

One of the reasons for the lack of results is that the complexity of the computations grows both with the number of operators and number of loops. One promising and alternative method to compute higher point  correlation functions  is the integrability based approach 
called hexagonalization
\cite{BKV,Hexagonalization,Eden1,HexagonalizationII,Eden2,HandlingI,HandlingII}.
 The main idea of the method is to cut the correlators into more fundamental objects, 
the hexagon form-factors, and 
then glue them back by inserting weighted complete sets of mirror particles states.  
Notice that both the hexagons and the others building blocks in this formalism are 
known at finite $g$ where $g$ is the coupling constant. 
The gluing process is the hard part of this approach and  
at the moment there is no efficient way of evaluating these corrections for a generic situation.
Moreover in some cases there are divergences
that require a still not fully understood regularization prescription \cite{WrappingOrder}. 
However, it has been recently shown \cite{Frank1,Frank2,KostovOctagon,KostovOctagon2} that by considering both very large operators in a particular polarization the glue back process simplifies and in the case of four operators the gluing process can be performed at finite coupling. 
This observable was dubbed the simplest four point function in $\mathcal{N}=4$ SYM
and in the integrability language it corresponds to an octagon square where 
by definition an octagon is the object obtained by gluing one length zero mirror edge of two hexagon form-factors with all the other edges having huge bridge lengths.
Giving the great success of the octagon, in this work 
we consider the next object in complexity which we call the simplest five point function 
or the decagon square. 
The gluing back process for the simplest five point function is considerably harder than 
the four point analogue. In this case, one has to glue two adjacent mirror edges and the mirror
particles in the different edges have a non trivial interaction. 
This implies that a new kind of integrability contribution shows up: the two-particle in different
edges contribution. This contribution was analysed for the first time at one-loop in 
\cite{HexagonalizationII} and its computation involves the mirror bound-state $S$-matrix 
which is a very complicated object. Due to the complexity of the decagon, we
will present the two-loop computation of it in this work. Nevertheless, along the way 
we have computed some components of the two-particle contributions at $l$ loop for arbitrary $l$ and
the full results will appear elsewhere.  
Recall that the simplest four-point function can be bootstrapped as it 
satisfies some all loop identities \cite{Frank2}. 
One of the goals of our two-loop computation is also to take the first steps in 
a possible analytical bootstrap approach to the decagon.  
The simplest five point result is given in terms of four different types of Feynman integrals with two of them being the well known one and two-loop ladder functions and the other two are a generalization of the ladder and pentaladder integral both involving the five external points. 
The new integrals are defined in (\ref{doubleboxmain}) and (\ref{pentaladdermain}). 

As mentioned above, the simplest four-point function involves very long operators and 
it is special polarized. One very interesting open problem is to 
study deformation of it, i.e. taking 
one of the bridges to have finite length. This problem also involves computing 
multi-particle integrability contributions and it is hard in general. 
In this work, we have computed 
some $l$-loop four-point functions using integrability by reducing our 
results from five  to four-points by identifying a pair of operators. 
Up to five-loops the results are not new and they have been computed
in \cite{3loopdata,45loopdata} by different techniques.  
In all cases where comparison is possible, we have got agreement with the
literature and this is a strong
check of our integrability calculations. 
Our approach for computing the integrability contributions was to generate 
power series and then fit the result against a basis of integrals taking into account 
the symmetries of the objects being computed. One drawback of this
method is that it is not possible to study four-point functions directly
because in order to generate the power series one has to resum parts of the integrand. 
Specifically, a four-point series does not truncate unless one performs a summation over one of the bound state indices 
in the integrability integrand.   
So, in this work, we have computed the integrability contributions for five points and then reduced 
the result to four by identifying a pair of points. This can be improved by a deeper understanding of the integrability integrand
and by performing some analytical integrations, see  \cite{deLeeuw:2019tdd,deLeeuw:2019qvz}
for interesting progress in this direction.  

This paper is organized as follows. Section \ref{sec:Integrability}  
has a brief review of the hexagonalization formalism and the multi-particle 
integrability contributions. In addition, the new two-loop expressions are presented and
their derivation explained. Our main result is in section \ref{planarcorrelationfunctions}
where we compute the simplest five point function at two-loops and a set of
four-point functions. The section  \ref{sec:IntegrabilityCalculation} has our 
conclusions and a list of possible continuations of this work. 
In the Appendices, we have a list of the necessary building blocks
and we also both explain how to compute the relevant Feynman integrals and derive 
differential equations for them.

\section{Integrability}\label{sec:Integrability}  

In this section, we briefly review the hexagonalization
procedure and its properties such as the flipping invariance and the coupling dependence of
several multi-particle configurations. 
We also explain how the two-particle
contributions at two-loop are computed and we give 
their results.        

\subsection{Review of hexagonalization} 

\begin{figure}[t]
\centering
\includegraphics[width=0.7\textwidth]{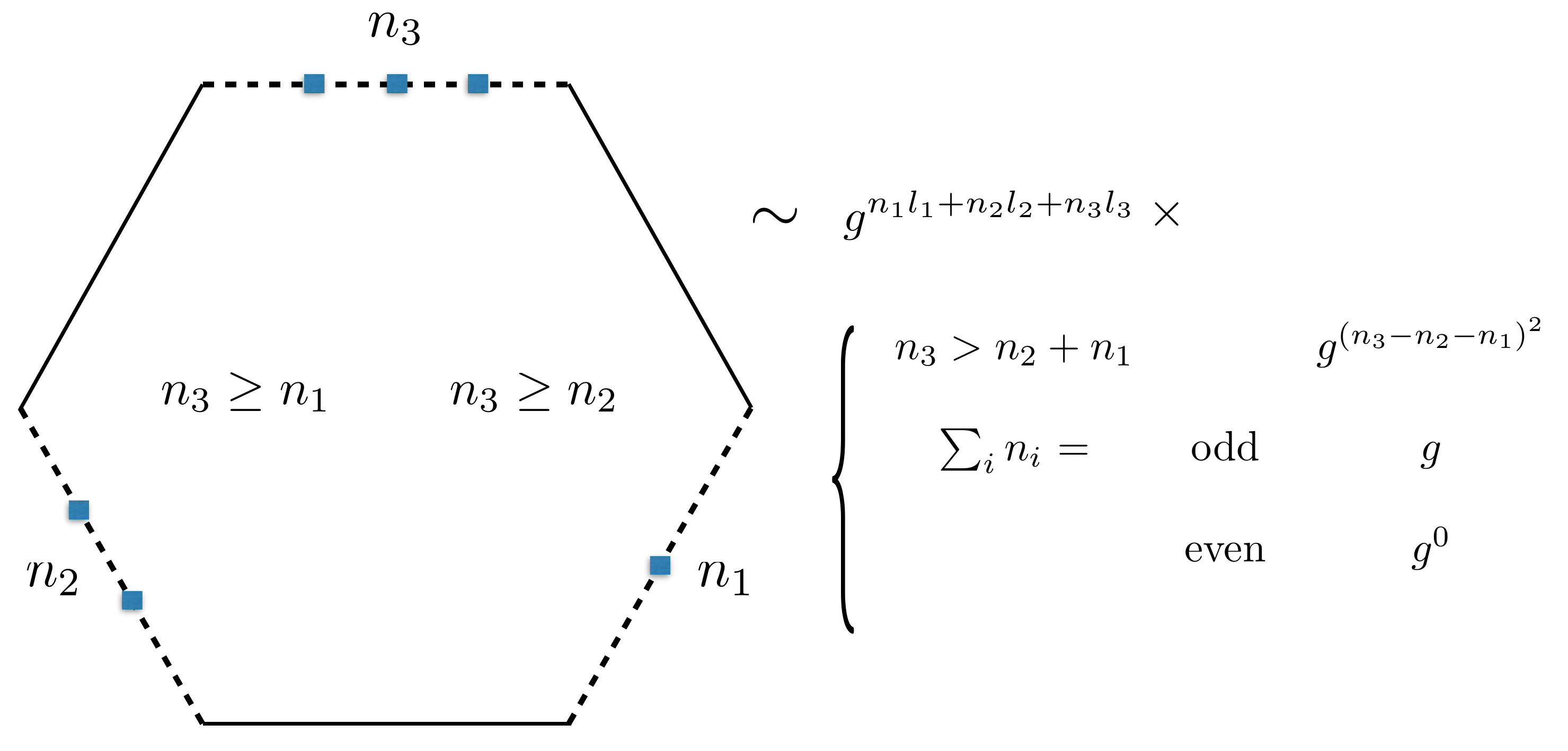}
\caption{This figure firstly appeared in  \cite{HandlingII} and it  is also discussed in 
\cite{Eden3}. The solid lines correspond to physical edges of the hexagon 
form-factor and the dashed lines
to mirror edges.  
The rule appearing in the figure 
enables one to estimative at each order in the coupling constant 
$g$ a configuration of mirror particles 
involving various hexagon form-factors kicks in. 
One only has to multiply the $g$ factor  obtained as in the 
figure for all the hexagons of a graph. 
Each excitation showed in the figure is properly normalized, i.e.  
it carries a square-root of the measure not explicitly indicated. The rule in the figure 
can be
derived by using the decoupling property of 
the hexagon form-factors and the $g$ dependence
of its dynamical part, see \cite{BKV}. The numbers $n_i$'s indicate
the number of mirror particles (elementary or bound-states) in each mirror edge. The $n_3$ is selected among the others by the constraints $n_3 \geq n_1$ and
$n_3 \geq n_2$.      
The $l_i$ indicates the bridge lengths of each mirror edge.}
\label{fig:ThePowerofg} 
\end{figure}

The hexagon form-factors $\mathcal{H}$ were firstly introduced in
 \cite{BKV} as an integrability based solution to the three-point
 function problem in $\mathcal{N}=4$ SYM. 
 The structure constants are written in this formalism 
 as a product of two hexagon form-factors and sums over both partitions of the physical external states and complete sets of mirror particles living in the 
so called mirror edges, we refer the reader to \cite{BKV} for details and explicitly 
expressions. 
The sum over mirror particles is equivalent to the insertion of a resolution of the identity and it is responsible for gluing the hexagons back to recover
the original object. 
The three-point function is an example of a more 
general procedure called hexagonalization where the hexagons are 
glued together to compute planar higher-point functions \cite{Hexagonalization, Eden1} 
and non-planar quantities \cite{HandlingI,HandlingII,Eden2}. In this work,
only the sphere topology will be considered, i.e. only 
planar correlators are going to be computed. In addition, all 
the external operators are going to be half-BPS operators and consequently there will be no physical
rapidities. This means that only the mirror edges of the hexagon form-factors will have particles, see figure 
\ref{fig:ThePowerofg}.   
Recall that a half-BPS operator $\mathcal{O}_L (x)$ 
is completely characterized by a null vector $y_I$, its position $x^{\mu}$ and its length $L$ and it is given by
\begin{equation}
\mathcal{O}_L(x) = {\rm{Tr}} \, ( (y \cdot \Phi(x))^L ) \, ,  
\end{equation}
where $\Phi^I(x)$ are the six scalars of $\mathcal{N}=4$ SYM.

The general procedure to compute a planar 
correlator using hexagonalization 
is to first list all tree-level graphs obtained 
by Wick contracting the operators and keeping 
only the connected planar ones and the 
disconnected ones that can be embedded in a
sphere. The propagators connecting the operators $i$ and $j$ are denoted by $d_{ij} = y_{ij}/x^2_{ij}$ and the number of equivalent ones 
(homotopically equivalent) connecting the operators are called the 
bridge lengths $l_{ij}$. 
Representing the equivalent propagators as a single line 
and drawing them using a double line
notation, one verifies that 
the propagators divide the sphere into faces. 
If the faces do not have an hexagonal shape (three physical and three mirror edges), one
can add additional lines (mirror lines) with zero bridge lengths connecting the 
operators in such way that all the surface is cut
into hexagons. In general, there are several ways 
of adding these extra lines but the final loop corrected result 
must not change; this is a consistency condition
of the formalism and it is called flip invariance. 
The loop corrections for each graph is obtained by promoting each hexagon to an hexagon
form-factor and by inserting a resolution of the identity in all mirror edges which means 
exciting mirror particles in those edges. Schematically, one has
\begin{equation}
\begin{aligned}
\langle \mathcal{O}_{L_1}(x_1) \ldots \mathcal{O}_{L_n}(x_n)    \rangle =  \hspace{10cm} \\
\mathcal{S} \cdot \left(  \sum_{{\rm{\underset{graphs}{tree-level} }}} 
\prod_{l_{ij}} (d_{ij})^{l_{ij}}  \right) 
\left( \sum_{i,j,k} \prod^{{\rm{hexagons}}}  
\mathcal{W}_{i,j,k}  \; \mathcal{H}_{\psi_{ij}, \psi_{jk}, \psi_{ki}} \right)  \, , 
\label{eq:GeneralFormula} 
\end{aligned}
\end{equation}
where $\mathcal{S}$ refers to stratification and it
is explained at length in \cite{HandlingII}.   
It is a procedure to properly take into account the graphs leaving in the boundary of the moduli space of 
the surfaces, for example the disconnected graphs 
for computing the planar correlators. 
These graphs can start to contribute at two-loop 
as the one-loop contribution was already proven to be zero in general in \cite{HandlingII}. The basic idea is that the effective genus of a graph can increase once virtual 
corrections are taken into account, this is similar to what happens in usual perturbation theory. The disconnected graphs are harder to compute as they 
involve more zero length bridges, fortunately they
do not play any role in this work because we only
compute special polarized five- and four-point functions and they not show up.       

In equation (\ref{eq:GeneralFormula}) 
all the elements appearing on the right hand side 
are known for any value of the coupling constant $g$.
The 
$\mathcal{H}_{\psi_{ij}, \psi_{jk}, \psi_{ki}}$ are the 
hexagon form-factors and the $\psi_{ij}$ are the set of mirror
particles leaving in the three mirror edges. 
Integration over the mirror particles rapidities is assumed. 
The  
$\mathcal{W}_{i,j,k}$ denotes the normalized weight factor depending on the flavour of the mirror-particles and 
on both the space-time and $R$-charge cross-ratios.  
Its explicitly expression will be given later. 

\begin{figure}[t]
\centering
\includegraphics[width=0.6\textwidth]{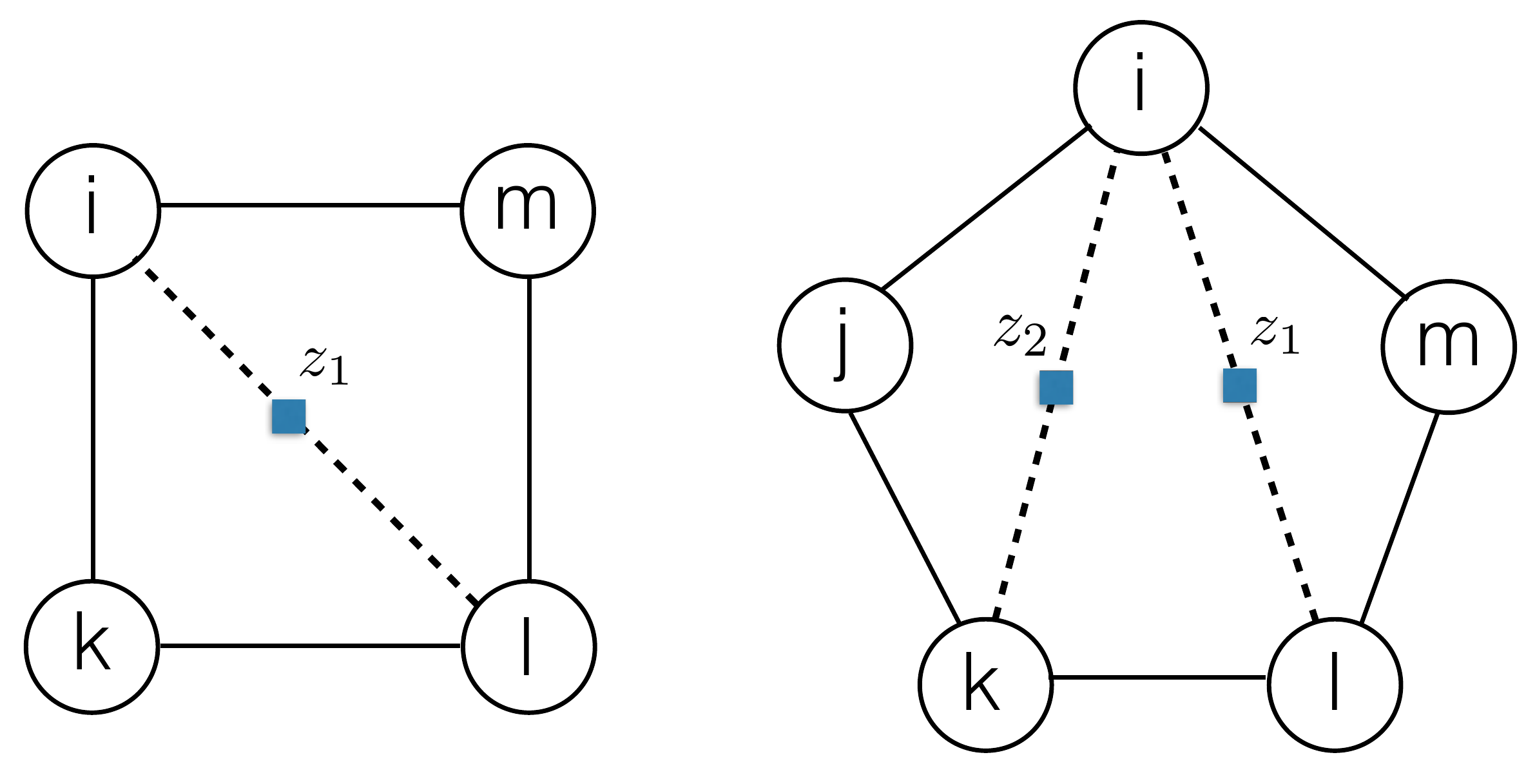}
\caption{The one- and two-particle length zero contributions.
The blue squares denote the mirror particles 
and the cross-ratios
$z_1$ and $z_2$ are defined in (\ref{eq:definitionofz1z2}).
The bridges connecting the operators (solid lines) 
have non-zero bridge-lengths. 
At two-loops, there are two relevant types of one-particle
contributions, one if the bridge length is zero as shown in the figure and the other if the brigde  has length one. Similarly, there are  three types of two-particle contributions depending on the bridge lengths, see (\ref{eq:types2}).}
\label{fig:TheDefiningCroosRatios} 
\end{figure}

To get the finite $g$ result of a correlator using hexagonalization is challenging as one has to resum
all the possible mirror particle configurations. 
As mentioned in the introduction this can be done 
only for very special correlators at the moment.   
Nevertheless, for a fixed order in $g$ only a finite set
of mirror particle configurations contribute because it
costs factors of $g$ to excite new particles. To estimate the order
in $g$ that a configuration of mirror particles kick in, one can use the rule described in the figure \ref{fig:ThePowerofg}. In this work, we are going to compute special 
polarized four- and five-point functions mostly at two-loops. In this case, only the one- and two-particle contributions showed in the figure  
\ref{fig:TheDefiningCroosRatios} where the
bridges lengths can be zero or one are going 
to be relevant. It is easy to verify, for example,
that a contribution of two-particles in the same mirror-edge kicks in at four loops or that a three-particle
contribution with two in the same edge and the other 
one in an adjacent edge kicks in at three-loops, see table \ref{table} for a more complete list. 
\begin{table}[h]
\begin{centering} 
\begin{tabular}{cc}
 $\{ n_1, n_2 \}$ & kick in at \\
 \hline
$\{ 1, 0 \}$ & $g^2$  \\
$\{ 2, 0 \}$ & $g^8$  \\
$\{ 1, 1 \}$ & $g^2$  \\
$\{ 2, 1 \}$  & $g^6$   \\
$\{ 3, 1 \}$  &  $g^{14}$ \\
$\{ 2, 2 \}$  &  $g^{8}$ \\
$\{ 3, 3 \}$  &  $g^{18}$ \
\end{tabular}
\caption{The order in $g$ that a configuration of multi-mirror particles leaving in one edge or adjacent edges kicks in. More specifically, consider the second graph of figure \ref{fig:TheDefiningCroosRatios}.
The number $n_1$ ($n_2$) correspond to the number of mirror particles in the edge denoted by $z_1$ ($z_2$) in that figure. The power of $g$ appearing in the table was computed by using the rule of figure 
\ref{fig:ThePowerofg}.  From the results in the table, we conclude that at 
two loop only the cases $\{n_{1}=1, n_{2} =0\}$  and $\{n_{1}=1, n_{2} =1\}$ contribute. 
 } 
\label{table}
\end{centering} 
\end{table}

The one-particle contribution at any loop and 
for any bridge length   
can be computed using the integrand 
given in equation (51) of \cite{Hexagonalization}. The two-particle
contribution at one-loop was firstly analysed in
\cite{HexagonalizationII} and its result at two-loop 
is computed in this paper. The relevant cross-ratios
to compute their contributions, 
see figure \ref{fig:TheDefiningCroosRatios}, are given  
by   
\begin{equation}
z_1 \bar{z}_1 = \frac{x_{im}^2 x_{kl}^2}{x_{ik}^2 x_{ml}^2} \, , \quad (1-z_1)(1-\bar{z}_1) = \frac{x^2_{il} x^2_{km}}{x^2_{ik} x^2_{lm}} \, , \quad 
z_2 \bar{z}_2 = \frac{x_{il}^2 x_{jk}^2}{x^2_{ij} x^2_{lk}} \, , \quad (1-z_2)(1-\bar{z}_2) = \frac{x_{ik}^2 x_{jl}^2}{x_{ij}^2 x_{kl}^2} \, .   
\label{eq:definitionofz1z2} 
\end{equation}
There are a similar set of $R$-charge cross-ratios
where the $x^2_{ij}$ are replaced by $y_{ij}$ and they will be denoted by $\{\alpha_1, \bar{\alpha}_1\}$ and $\{\alpha_2,\bar{\alpha}_2\} $. In what follows, we will always work with a restricted kinematics, i.e. we are going to consider that all the space-time points and
polarizations are restricted to a given plane.
The reason for this restriction is that the integrability calculation is easier in this case as the
general weight factor for gluing the hexagons when the operators are out of the plane is more complicated. In this restricted kinematics the fifth cross-ratio is a function of
the ones introduced before and it is given by 
\begin{equation}
\frac{x^2_{ik} x^2_{mj} }{x^2_{im} x^2_{kj}} = \frac{(1-z_1+ z_1 z_2)(1-\bar{z}_1+ \bar{z}_1 \bar{z}_2)}{z_1 \bar{z}_1 z_2 \bar{z}_2}  \, , 
\end{equation} 
and similarly for the fifth $R$-charge cross-ratio.

In this final part of the subsection, we are going
for the convenience of the reader to compile some
of the results known in the literature 
\cite{Hexagonalization,HexagonalizationII} about the 
mirror particle contributions that are going to be used 
later. The new results at two-loop
are going to be presented with further explanations
in the next subsection. In our conventions,
the mirror particles contributions are going to be denoted by 
\begin{equation}
\mathcal{M}^{(L)}_{n,\{l_1\}}(z_1) \, ,  \quad {\rm{and}} \quad 
\mathcal{M}^{(L)}_{n,\{l_1,l_2\}}(z_1, z_2) \, , 
\label{eq:Representation}
\end{equation}
and the arguments are read anti-clockwise with respect to its graphic 
representation, see figure \ref{fig:TheDefiningCroosRatios}. In (\ref{eq:Representation}), $L$ denotes the loop 
order, $n$ is the number of particles involved
and $\{l_1,l_2\}$ has information about the bridge lengths of the mirror edges. 
Note that the mirror particle contributions also depend on the $R$-charge cross-ratios 
$\{ \alpha_1, \alpha_2 \}$,
but we have omitted them as they can be deduced unambiguously from 
the dependence of the space-time cross-ratios.

All the one-particle contributions can be written 
in terms of the following function   
\begin{equation}
m^{(L)} (z) \equiv g^{2 L} \frac{(z+ \bar{z}) - (\alpha + \bar{\alpha})}{2} F^{(L)} (z, \bar{z})  \, ,  \quad  
{\rm{with}} \quad g^2 = \frac{\lambda}{16 \pi^2} \, ,
\end{equation}
and $\lambda$ is the t'Hooft coupling. 
It is clear that $m^{(L)}(z)$ depends on $z$ and $\alpha$, but again we are excluding $\alpha$ from the list of arguments.  
Moreover,  
\begin{equation}
F^{(L)}(z, \bar{z} ) = \frac{1}{z - \bar{z}} 
\left[ \sum_{k=0}^L \frac{(-1)^k (2L-k)!}{L!(L-k)!k!} 
{\rm{log}}^{k} (z \bar{z}) ({\rm{Li}}_{2L -k} (z) 
-{\rm{Li}}_{2L -k} (\bar{z}) )
\right] \, .
\label{eq:LadderIntegralResult}
\end{equation}                                                                                                                                                                                 
The function $F^{(L)}(z, \bar{z} )$ given above is related to the so called 
ladder integrals \cite{ladder}. For example,
\begin{equation}
F^{(1)}(z, \bar{z}) = \frac{x_{13}^2 x^2_{24}}{\pi^2} \int \frac{d^4 x_5}{x_{15}^2 x^2_{25} x^2_{35} x^2_{45}} \, , \quad 
F^{(2)}(z, \bar{z}) = \frac{x^2_{14}x_{13}^2 x^2_{24}}{(\pi^2)^2} \int \frac{d^4 x_5 d^4 x_6}{x_{15}^2 x^2_{25} x^2_{45} x^2_{56} x^2_{16} x^2_{36} x^2_{46}} \, , 
\end{equation}     
with                            
\begin{equation}
z \bar{z} = \frac{x_{12}^2 x_{34}^2}{x^2_{13} x^2_{24}} \, , \quad  \quad (1-z)(1-\bar{z}) = \frac{x^2_{23} x^2_{14}}{x^2_{13} x^2_{24}} \, . 
\label{eq:Crosszzb} 
\end{equation} 
The one-loop length zero one-particle contribution was originally computed
in \cite{Hexagonalization} and it is given by 
\begin{equation}
\mathcal{M}_{1,\{0 \} }^{(1)}(z)  
= m^{(1)}(z) + m^{(1)}(z^{-1}) \, . 
\end{equation}                  
Using the integrand appearing in that same paper, it is not dificult to compute other one-particle contributions by doing the integration by residues and explicitly performing the summation
over bound-states. 
In fact, there is a closed expression for it, see \cite{Frank1} and (\ref{oneparticlegeneralformula}).
For this work we will need the following 
two loop and three loop contributions
\begin{equation}
\begin{aligned}
&\mathcal{M}_{1,\{1\}}^{(2)}(z)  
= m^{(2)}(z) + m^{(2)}(z^{-1}) 
\, , \quad \mathcal{M}_{1,\{0\}}^{(2)}(z) =-2 \, \mathcal{M}_{1,\{1\}}^{(2)}(z) \, .  \\
&\mathcal{M}_{1,\{2\}}^{(3)}(z)  
= m^{(3)}(z) + m^{(3)}(z^{-1}) \, , \quad
\mathcal{M}_{1,\{1\}}^{(3)}(z) =-4 \, \mathcal{M}_{1,\{2\}}^{(3)}(z) \, , \\
&\mathcal{M}_{1,\{0\}}^{(3)}(z) =6 \, \mathcal{M}_{1,\{2\}}^{(3)}(z) \, .
\end{aligned} 
\end{equation}   
Notice that all the expressions above are 
invariant under $z \rightarrow 1/z$. This invariance is manifest when writing them in terms of the functions $m^{(L)}(z)$ and it follows because of the important property of the ladder integrals
\begin{equation}
F^{(L)}(1/z, 1/\bar{z}) = z \bar{z}  \,  F^{(L)}(z \bar{z} ) \, .
 \end{equation} 
 This invariance of the one-particle contribution is called flip invariance. It corresponds to 
 the invariance of the position of  the additional length zero bridge in the graphs
 or the invariance under different tesselations. 
 Concretely, if one moves the dashed line of the first graph of the figure \ref{fig:TheDefiningCroosRatios} to connect the operators 
 $\mathcal{O}_m$ and $\mathcal{O}_k$ instead of the operators $\mathcal{O}_i$ and 
 $\mathcal{O}_l$ the relevant cross-ratio change as $z_1 \rightarrow 1/z_1$.

 The two-particle contribution appearing in the figure 
 \ref{fig:TheDefiningCroosRatios} was computed in \cite{HexagonalizationII} at one-loop. Its calculation will be reviewed and  extended in the next subsection. 
The result at one-loop is   
\begin{equation}
\begin{aligned}
\mathcal{M}^{(1)}_{2; \{ 0,0 \} }
(z_1,z_2) =
& -m^{(1)}(z_1) -m^{(1)}(z_2^{-1}) \\ 
& + m^{(1)} \left( \frac{z_1-1}{z_1 z_2} \right)  
+ m^{(1)} \left(\frac{1-z_1+z_1 z_2}{z_2} \right) 
+m^{(1)}(z_1(1-z_2)) \, . 
\label{eq:Onelooptwoparticle}
\end{aligned} 
\end{equation}
\begin{figure}[t]
\centering
\includegraphics[width=0.7\textwidth]{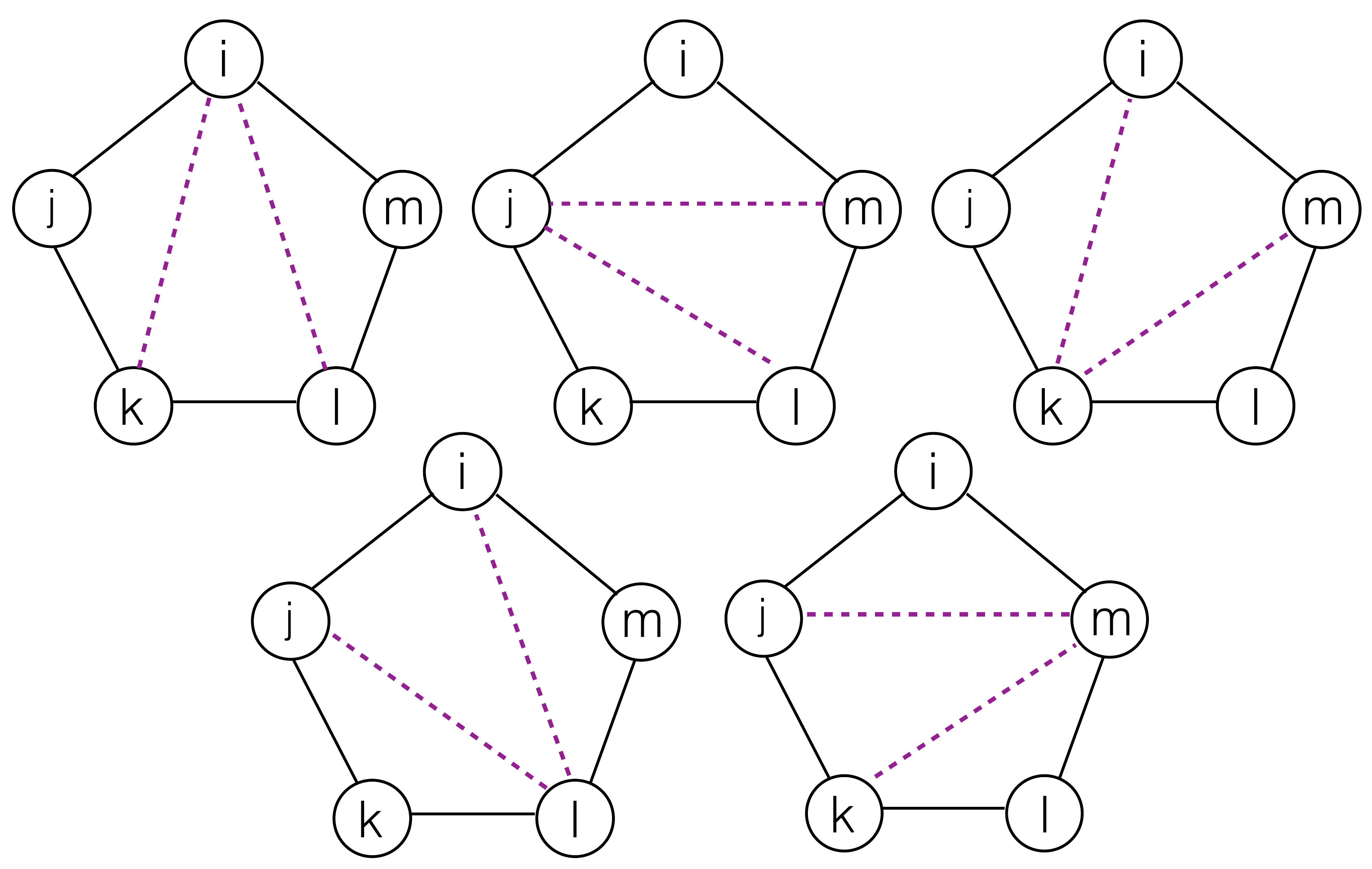}
\caption{The five different ways of cutting the inside of a decagon. The outside part can be cut similarly in five different ways. If the solid lines have bridge length greater or equal to two, at one- and two-loops only the one-particle and two-particle contributions of figure 
\ref{fig:TheDefiningCroosRatios} have to be taken into account
for its computation. The result of the computation has to be the same for all the cuts, {\em{i.e.}} the result is 
rotation invariant, for example, by doing $i\rightarrow m,m \rightarrow l \dots j\rightarrow i$, the
result does not change.}
\label{fig:TheFlippingFive} 
\end{figure}

An interesting object to compute is the decagon showed in the figure \ref{fig:TheFlippingFive}
with five different tesselations. It appears as a contributing diagram in several five-point functions. At one-loop, it is given
by a sum of two one-particle contributions and a two-particle contribution. This sum is equal to        
\begin{equation}
\begin{aligned}
\mathcal{M}^{(1)}_{1,\{ 0 \}}(z_1, \alpha_1)
&+ \mathcal{M}^{(1)}_{1,\{ 0 \} }(z_2, \alpha_2)
+ \mathcal{M}^{(1)}_{2; \{0,0\}} 
(z_1, z_2, \alpha_1, \alpha_2) = \\
 m^{(1)}(z_1^{-1}) &+  m^{(1)}(z_2) + m^{(1)} \left( \frac{z_1-1}{z_1 z_2} \right) + m^{(1)} \left(\frac{1-z_1+z_1 z_2}{z_2} \right) 
+ m^{(1)}(z_1(1-z_2)) \, . 
\label{oneloopdecagon}
\end{aligned} 
\end{equation} 
The decagon is flip invariant or in other words the  result is the same if we cut it in any of the different ways showed in the figure \ref{fig:TheFlippingFive}. The right hand side of the expression (\ref{oneloopdecagon}) is manifestly flip invariant because
if one performs rotations on the figure (for example
 $i\rightarrow m,m \rightarrow l \dots j\rightarrow i$)
 the cross-ratios appearing as arguments of the functions $m^{(1)}$ are mapped into themselves.  
As we will see, at two-loop we have similarly that the right hand side 
is given by one function evaluated at five different points. 

\subsection{Two-particle contributions at two-loop} 
\begin{figure}[t]
\centering
\includegraphics[width=0.6\textwidth]{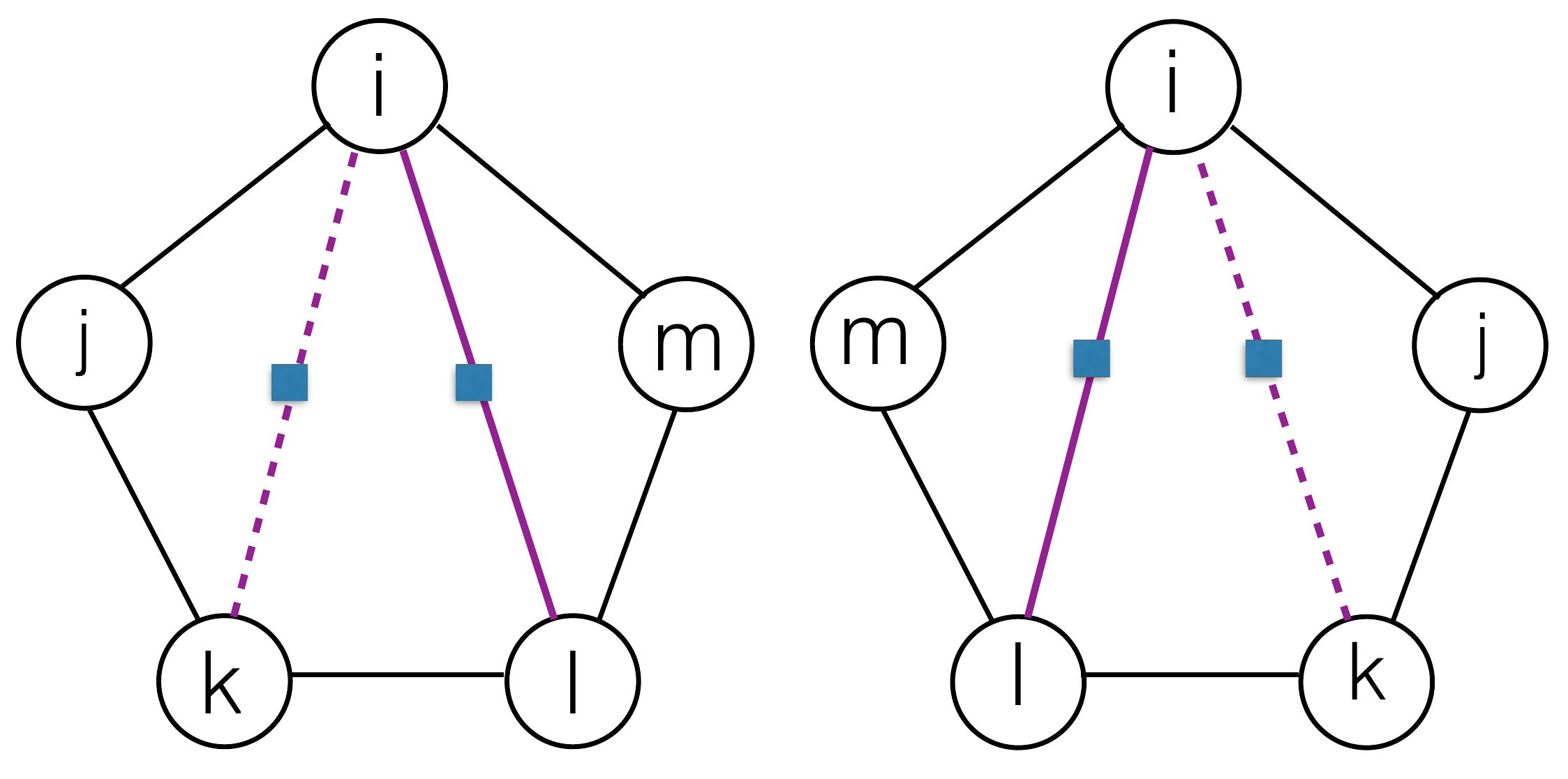}
\caption{
The two-particle contributions $\mathcal{M}^{(2)}_{2,\{1,0 \} }$ and 
$\mathcal{M}^{(2)}_{2,\{0,1 \} }$  with nonzero bridges lengths. In the figure, the solid purple line 
has bridge length one and the dashed purple line has bridge length zero. 
The blue squares denote the mirror particles
Notice that the two figures represents the same 
graph and they are related by a parity transformation.
The equality of the graphs implies the all loop parity relation $\mathcal{M}^{(L)}_{2;\{1,0\}}(z_1 ,z_2) = \mathcal{M}^{(L)}_{2;\{0,1\}}(z_2^{-1},z_1^{-1})$.} 
\label{fig:ParityInvariance} 
\end{figure}
At two-loop there is
three types of two-particle contributions depending on the length of the bridges involved 
(the cross-ratios are defined in (\ref{eq:definitionofz1z2})):
\begin{equation}
\mathcal{M}^{(2)}_{2,\{0,0 \} }(z_1, z_2) \, , \quad
\mathcal{M}^{(2)}_{2,\{1,0 \} }(z_1, z_2)  \, , \quad
\mathcal{M}^{(2)}_{2,\{0,1 \} }(z_1, z_2) \, .   
\label{eq:types2}
\end{equation}    
The first one on the list above corresponds  
to the second graph of figure  \ref{fig:TheDefiningCroosRatios} and it involves two bridges of length zero. The other two 
involves both a bridge of length one and a bridge of length zero,
see figure \ref{fig:ParityInvariance}. The same figure 
shows a graph before and after a parity transformation and the invariance 
of the result implies the following relation for any loop order or any $L$
\begin{equation}
\mathcal{M}^{(L)}_{2;\{1,0\}}(z_1 ,z_2) = \mathcal{M}^{(L)}_{2;\{0,1\}}(z_2^{-1},z_1^{-1}) \, .
\label{eq:ParityInvariance}
\end{equation}
A similar reasoning implies  
\begin{equation}
\mathcal{M}^{(L)}_{2,\{0,0 \} }(z_1, z_2) =\mathcal{M}^{(L)}_{2,\{0,0 \} }(z_2^{-1}, z_1^{-1}) \, .   
\label{eq:ParityLengthZero} 
\end{equation}

\begin{figure}[t]
\centering
\includegraphics[width=0.7\textwidth]{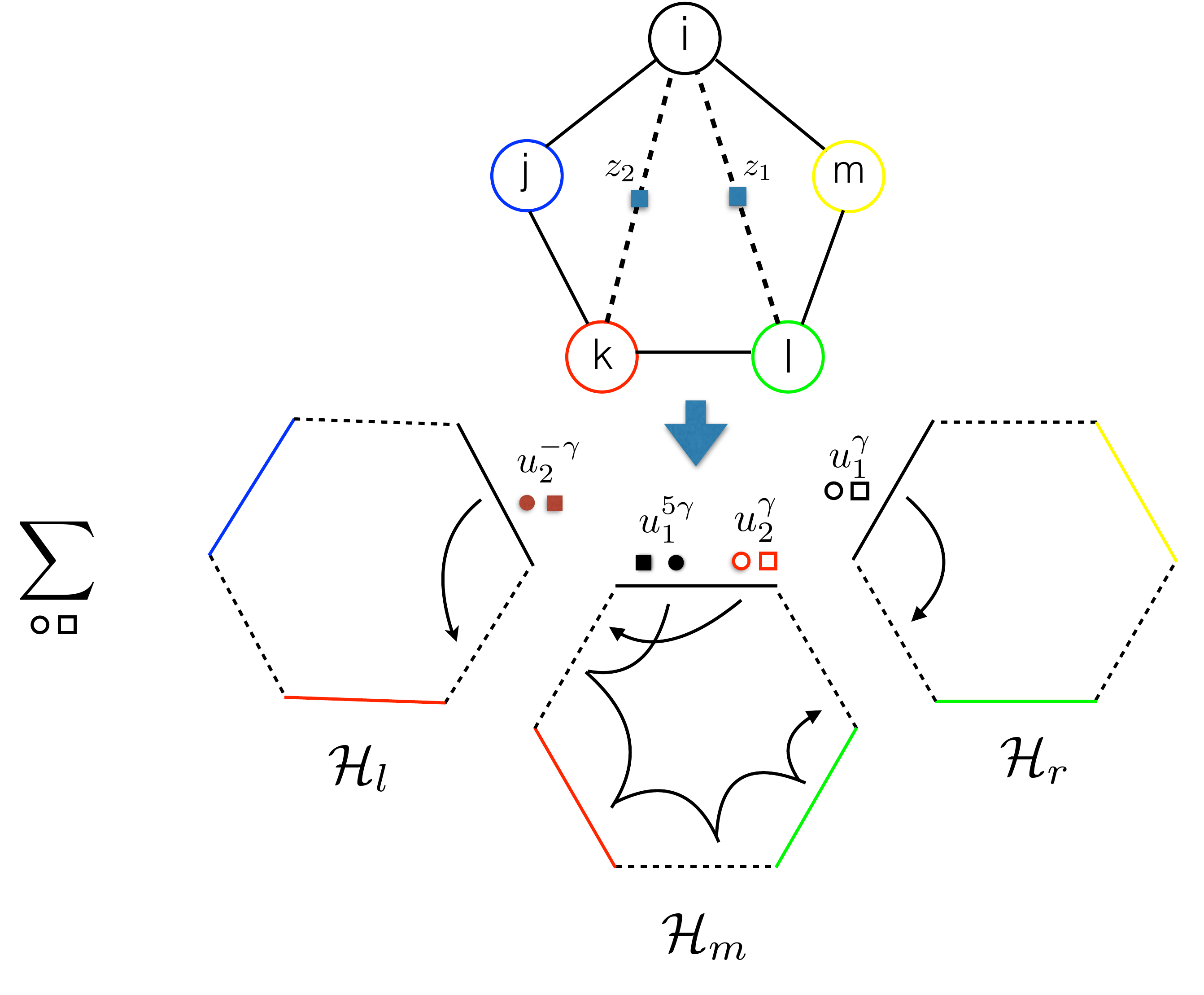}
\caption{The computation of the two-particle contribution. At one-loop, it was computed in \cite{HexagonalizationII}. The calculation involves three hexagons. The left ($\mathcal{H}_l$)  and right
($\mathcal{H}_r$) hexagons contains only one mirror particle and they contribute only with a possible sign.       The middle ($\mathcal{H}_m$) hexagon has two particles and its computation is non-trivial as it involves a dynamical factor plus the mirror-mirror bound state $S$-matrix. The way of contracting the indices of the mirror particles 
is shown in the figure \ref{fig:Smatrix}.    
To glue the hexagons one sums over a weighted  complete set
of mirror states and the empty and solid circles and squares  of the same color correspond to
a pair of conjugate indices.}
\label{fig:TheHexagons2} 
\end{figure}
The computation of the two-particle contributions involves three hexagons, see figure \ref{fig:TheHexagons2}. At one-loop, it was computed
in \cite{HexagonalizationII}. The computation at two-loops follows the same steps 
of the one-loop calculation described in that paper. One only needs to
expand all the building blocks at higher order in $g$ or
add a bridge of length one to the integrand. 
The calculation involves three hexagon form-factors 
(left, middle and right) denoted by
$\mathcal{H}_l$, $\mathcal{H}_m$ and $\mathcal{H}_r$ respectively in the figure 
\ref{fig:TheHexagons2}. To glue the hexagons  
one sum over a weighted complete set
of states (fundamental and bound states). We need to glue two edges so there will be a double sum over bound states and a double integral over the rapidities
$u_1$ (particles of the first edge) and $u_2$ (particles of the second edge) of the mirror particles. One has after collecting all the
ingredients 
(see the appendix \ref{weakexpansion}): 
\begin{equation}
\begin{aligned} 
\mathcal{M}_{2, \{ l_1, l_2 \}} (z_1, z_2) 
=  \frac{1}{2} \sum_{{\rm{dressings = +, -}}} \int \frac{d u_1}{2 \pi} \frac{d u_2}{ 2 \pi} 
\sum_{a=1}^{\infty} \sum_{b=1}^{\infty} 
\mu_{a}(u_1^{\gamma}) 
\mu_{b}(u_2^{\gamma}) e^{- \tilde{E}_a(u_1) l_1} 
e^{- \tilde{E}_b(u_2) l_2} \\
\sum_{I, J} \mathcal{W}_{\{ b, J \} }(u_2^{\gamma} )   
\, \mathcal{W}_{\{ a, I \} }(u_1^{\gamma} ) \, \mathcal{H}_{l} [ \, \bar{\chi}_{\{b,J\}} (u_2^{- \gamma} )\,]   
\, \mathcal{H}_{m} [  
\chi_{\{b,J\}} (u_2^{\gamma} ) \bar{\chi}_{\{a,I\}} (u_1^{- \gamma} ) ] 
\mathcal{H}_{r} [ \chi_{\{a,I\}} (u_1^{\gamma} )] \;  \, .
\label{eq:Formula2twoparticle} 
\end{aligned}
\end{equation}                
In the expression above $\mu_a$ is the measure 
or a normalization factor and the $\gamma$ over the rapidities are mirror transformations. 
The $\tilde{E}$ are the
mirror particle energies and they are multiplied
by the bridge lengths so they correspond to damping factors.
The indices $a$ and $b$ are bound state indices 
and $I$ and $J$ are sums over complete set of 
states.
More specifically, each magnon $\chi_{\{b\}}$ is schematically of the form 
\begin{equation}
\chi_{\{b\} c \dot{c}} = | \mathcal{X}_{c} \rangle_b \otimes | \mathcal{X}_{\dot{c}}  \rangle_b \, ,   
\end{equation}
where $c$ and $\dot{c}$ are sets of fundamental indices
of the BMN symmetry group $\mathfrak{psu}(2|2) \otimes \mathfrak{psu}(2|2)$
and $| \mathcal{X}_c \rangle_b$ is a basis 
for the $b$-th anti-symmetric representation which has dimension $4 b$. This means that the index $I$ in (\ref{eq:Formula2twoparticle}) labels $(4 a)^2$ states and similarly for $J$.
The basis states are built from the fundamental fields
$\phi_a$ (bosonic) and $\psi_{\alpha}$ (fermionic) and they are  
listed in the Appendix B of \cite{HexagonalizationII}.
Note that the bar over $\chi$ means conjugation of all 
the indices of the fields $\phi_1  \leftrightarrow \phi_2$
and $\psi_1  \leftrightarrow \psi_2$. 
The foremost sum in the expression
(\ref{eq:Formula2twoparticle}) is over the dressings,
in other words, the naive basis of the bound states  
has to be modified by inserting some $Z$-markers in two different ways denoted plus and minus (the factor of half is because one
has to take the average, see \cite{deLeeuw:2019tdd,deLeeuw:2019qvz} 
for a recent 
discussion about $Z$-markers).  The rule for the plus dressing is the following 
\begin{equation}
{\rm{dressing}} \, \, + \, : \, 
 \psi_{\alpha} \rightarrow \psi_{\alpha}  \,  ,  \; 
\psi_{\dot{\alpha}} \rightarrow \psi_{\dot{\alpha}} \, , \; 
\{ \phi_{1}  \, , \phi_{\dot{2}} \}  \rightarrow  
Z^{\frac{1}{2}} \{ \phi_{1}  \, , \phi_{\dot{2}} \}  \, , \; 
\{ \phi_{\dot{1}}  \, , \phi_{2} \}  \rightarrow  
Z^{-\frac{1}{2}} \{ \phi_{\dot{1}}  \, , \phi_{2} \}   \, . 
\end{equation}
and for the dressing $-$, one needs to change the sign of the exponents  of all the $Z$-markers above. 
One consequence of dressing the basis  
is that the $Z$-markers will appear inside the hexagon form-factors in the expression 
(\ref{eq:Formula2twoparticle}). They can them be 
moved and removed using the rules given in the Appendix C of \cite{BKV}. Another consequence is that they 
give contributions to the weight factors $\mathcal{W}$,
which is given by 
\begin{equation}
 \mathcal{W}^{\pm}_{\{ a, I \} }(u_i^{\gamma} )  
 = e^{-2 i \tilde{p}_a(u_i) {\rm{log}} |z_i| }  \, 
 e^{i L \phi_i }  \, e^{i R (\theta_i \pm \varphi_i)} \, ,
\label{weightfactor} 
\end{equation}
where the $\pm$ refers to the two dressings, $\tilde{p}_a(u_i)$ is the mirror momentum and the angles are 
defined in terms of the cross-ratios (\ref{eq:definitionofz1z2}) as follows 
\begin{equation}
e^{i \phi_i} = \sqrt{\frac{z_i}{\bar{z}_i}}  \, , \quad 
e^{i \theta_i} = \sqrt{\frac{\alpha_i}{\bar{\alpha}_i}} \, , \quad 
e^{i \varphi} = \sqrt{\frac{ \alpha_i \bar{\alpha}_i }{ z_i \bar{z}_i }} \, .
\end{equation}
Finally, $L$ and $R$ in (\ref{weightfactor}) are combinations of diagonal Lorentz and $R$-charge generators given by
\begin{equation}
L = \frac{1}{2} ( L^{1}_{\; 1} -  L^{2}_{\; 2}  -  L^{\dot{1}}_{\; \dot{1}} +L^{\dot{2}}_{\; \dot{2}}) \, , \quad 
R = \frac{1}{2} ( R^{1}_{\; 1} -  R^{2}_{\; 2}  -  R^{\dot{1}}_{\; \dot{1}} +R^{\dot{2}}_{\; \dot{2}}) \, . 
\end{equation}
Notice that the weight factors  
act on the particles corresponding to its argument and 
all the basis elements for the bound states 
are eigenstates of the above generators (the action of them on the particles 
are canonical).                          
\begin{figure}[t]
\centering
\includegraphics[width=0.5\textwidth]{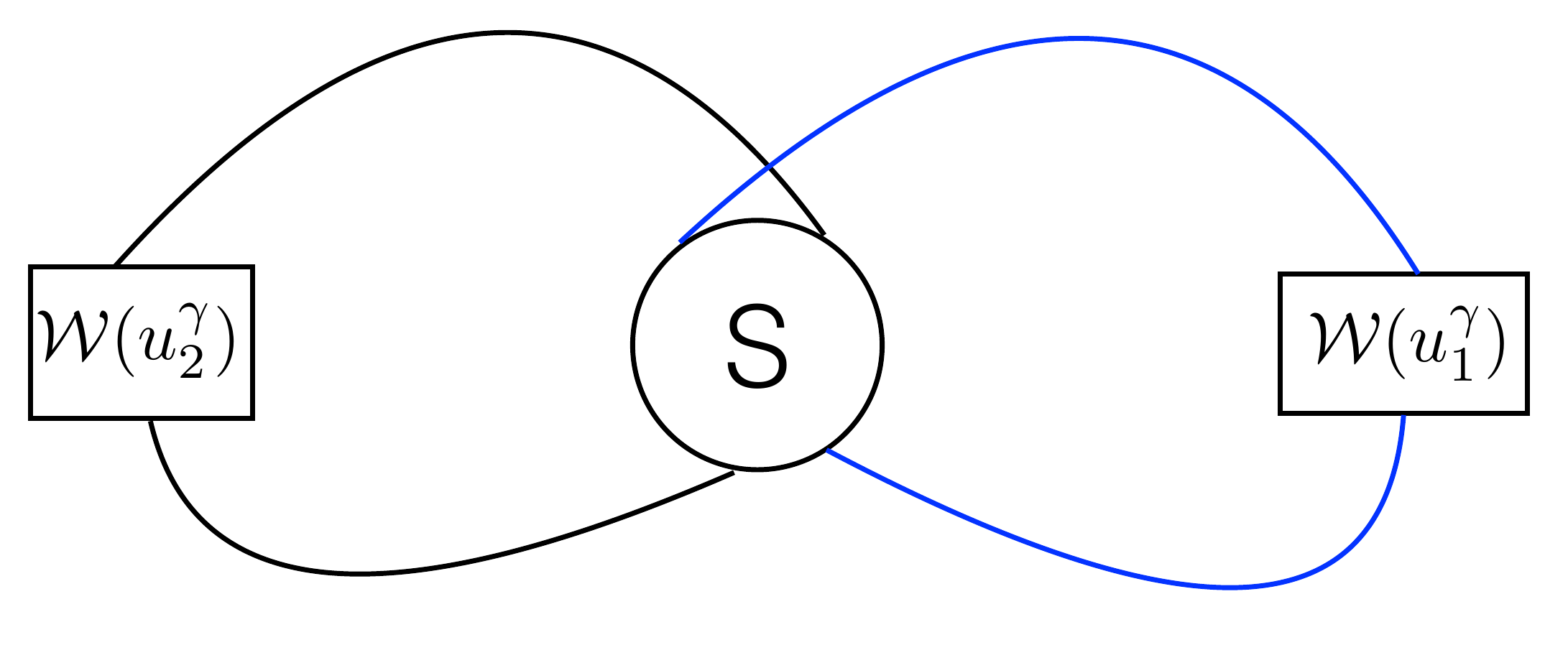}
\caption{The matrix part of the two-particle calculation of (\ref{eq:Formula2twoparticle}). The 
left ($\mathcal{H}_l$) and right ($\mathcal{H}_r$) hexagon form-factors have only one-particle (fundamental or bound-state) and 
they are easily evaluated. They give a non-zero
contribution only for $\chi$'s where the dotted indices are conjugate to the undotted indices ($1 \leftrightarrow 2$) as the one-particle 
hexagon form-factor is a product of $\epsilon^{A B}$.  
 This forces the scattering 
between the particles in the middle hexagon $\mathcal{H}_m$ to be diagonal as shown in the figure where $S$ means mirror bound state $S$-matrix. The $\mathcal{W}$ are weight factors and they 
act on the particles corresponding to its argument.}
\label{fig:Smatrix} 
\end{figure}

To compute (\ref{eq:Formula2twoparticle}) 
we need to evaluate three hexagons form-factors. The hexagons have a dynamical scalar factor and a matrix part that boils down to a
product of $S$-matrices. The hexagons $\mathcal{H}_l$ and $\mathcal{H}_r$ have only 
one-particle and they evaluate to zero, one or minus one. The non-zero cases occurs when the undotted 
indices are conjugate to the dotted indices, i.e. 
only when one excites the so called tranversal excitations (the scalars $Y^{1 \dot{2}}$, $\bar{Y}^{2 \dot{1}}$ or the derivatives $D^{1 \dot{2}}$, $\bar{D}^{2 \dot{1}}$) or fused products of it, see \cite{BKV}.  The middle hexagon $(\mathcal{H}_m)$ has two-particles and it is non-trivial. In order to compute it we are going by convenience to analytically continue the rapity $u_1$ from  $u_1^{-\gamma}$ to $u_1^{5 \gamma}$ as shown in figure 
\ref{fig:TheHexagons2}. 
The matrix part of the middle hexagon is the mirror bound
state $S$-matrix computed in \cite{HexagonalizationII} by adapting the previous calculations of the physical bound state $S$-matrix
of \cite{physicalSmatrix}. The bound state $S$-matrix 
is block diagonal and its blocks can be organized
in three distinct cases. The fact that the left and right hexagons are non-zero only for the tranverse excitations 
forces the scattering in the middle hexagon to be diagonal and the matrix part of it is shown in figure
\ref{fig:Smatrix}.  The full integrand after 
the evaluation of the hexagon form-factors is very length and it can be found at any loop order for the case with two length 
zeros in the Appendix D of  \cite{HexagonalizationII} .   
The general case for non-zero bridge lengths only amounts to include 
the damping factors appearing in (\ref{eq:Formula2twoparticle}).  
In this paper, we are going to evaluate the two-loop integrals by expanding the building blocks 
of the integrand (\ref{eq:Formula2twoparticle})  to order $g^4$. Notice that at this order it is only
possible to have bridge lengths zero and one becuase 
\begin{equation}
e^{- \tilde{E}_a(u_1) l_1}  \, \sim \, \mathcal{O}(g^{2 l_1} ) \, . 
\end{equation}  

Expanding the momentum factors of the two length zero integrand appearing from removing
the $Z$-markers one verifies that new $R$-charge 
structures appears at two-loop\footnote{At the moment we do not have 
an explanation for the $Z$-markers prescription. It is possible that they are 
related to a kind of spin-structure. Moreover, we are using an hybrid convention for the generators and the fermionic transformations of the particles involve powers of $Z$-markers, thus the $Z$-marker prescription is related somehow to supersymmetry.
Notice that they are responsible for removing the square root cuts of all integrability integrands and in this sense the prescription is almost unique \cite{deLeeuw:2019tdd,deLeeuw:2019qvz}. From the relation to perturbative calculations it is possible to understand qualitatively the increasing of $R$-charge structures of the multi-particle integrability contribution. At one-loop there is a map between the integrability contributions and the supergluon exchange in the $\mathcal{N}=2$ formulation
\cite{Eden2}. At two-loop the computation involves multi gluon exchanges and it is more complicated.}.   
Specifically, one has that the two-particle length zero contribution
involves the following structures kicking in at
the following loop order
\begin{equation}
\begin{aligned}
g^2 :   \{ \alpha_1 \alpha_2, \frac{1}{\alpha_1 \alpha_2}, 
\alpha_1, \frac{1}{\alpha_2}, \frac{\alpha_1}{\alpha_2}
\}  \, , \quad 
g^4:  \,\{ \frac{1}{\alpha_1} , \alpha_2 \} \, , \quad 
g^6:  \{ \frac{\alpha_2}{\alpha_1}  \} \, .
\end{aligned} 
\end{equation}
Notice that no new structure appears at four-loops and beyond. In addition, the result is symmetric under
the exchange $\alpha \leftrightarrow \bar{\alpha}$
so the structures above always appears in a combination with the bar ones.     

The integrand involves various sums and
it is complicated to analytically evaluate the integral in general 
without further simplifications or a more deep understanding of it. The only known 
complet analytically 
evaluation of a two-particle contribution is the simpler case of the tree-level propagator contribution for fishnet theories appearing in \cite{FishnetHexagons}.  
It should be possible to extend these 
tecniques to more complicated integrals or
greatly simplifies the integrand, we hope 
to address these questions in the future and some recent progresses in evaluating 
these integrals analytically can be found in \cite{deLeeuw:2019tdd,deLeeuw:2019qvz}.
In this work, we only generate power series from integrability and fit the results with a basis of two-loop integrals that we evaluate as described in the Appendix \ref{Feynmanintegrals}. After the fitting, we have double 
checked the result by computing the differential equations obeyed by the integrals and applying
them to the integrability series. Some 
care is needed in deriving the differential equations
because we are considering a restricted kinematics where the five-points are in a plane, see also the 
Appendix \ref{Feynmanintegrals}. The basis of integrals we have used 
consists of the ladder integral $F^{(2)}(z, \bar{z})$ given in (\ref{eq:LadderIntegralResult}), the double
box integral\footnote{This integral for five-points at generic positions depends on five cross-ratios instead
of four. The result for the general case is known and it was obtained by Matthias Wilhelm in \cite{MatthiasPrivate}. 
In order to reproduce Matthias's result from integrability,  it is necessary to use an out of plane weight factor.}    
\begin{align}
L_{1,24,35}=\int \frac{d^4x_6d^4x_7}{x_{16}^2x_{27}^2x_{36}^2x_{67}^2x_{47}^2x_{56}^2x_{17}^2}= \frac{L\left(\frac{x_{12}^2x_{34}^2}{x_{14}^2x_{23}^2},\frac{x_{13}^2x_{24}^2}{x_{14}^2x_{23}^2},\frac{x_{15}^2x_{34}^2}{x_{14}^2x_{35}^2},\frac{x_{13}^2x_{45}^2}{x_{14}^2x_{35}^2}\right)}{x_{13}^2x_{14}^2x_{25}^2} \, , 
\label{doubleboxmain}
\end{align}
and the pentaladder integral 
\begin{align}
P_{1,34,25}=\int\frac{d^4x_6d^4x_{7}\,x_{17}^2}{(x_{16}^2x_{46}^2x_{36}^2)x_{67}^2(x_{47}^2x_{37}^2x_{27}^2x_{57}^2)}= \frac{P\left(\frac{x_{12}^2x_{34}^2}{x_{14}^2x_{23}^2},\frac{x_{13}^2x_{24}^2}{x_{14}^2x_{23}^2},\frac{x_{15}^2x_{34}^2}{x_{14}^2x_{35}^2},\frac{x_{13}^2x_{45}^2}{x_{14}^2x_{35}^2}\right)}{(x_{34}^2)^2x_{25}^2} \, .
\label{pentaladdermain}
\end{align}
Notice that the first subindex is special in this two integrals. Since for $L$ it has conformal weight two and is connected to both integration points and for $P$ it appears both in the numerator and in the denominator. 

Our result for the two-particle contribution at two-loop  involving 
bridges of lengths zero and one is   ($z_1$ and $z_2$ are defined in (\ref{eq:definitionofz1z2})
and we are using the conventions of the figure \ref{fig:TheDefiningCroosRatios})  
\begin{equation}
\begin{aligned} 
 \mathcal{M}^{(2)}_{2,\{a,b\}}(z_1,z_2) &= - f(z_1)K^{1}_{\{a,b\}}(z_1,z_2) - f(z_2^{-1})K^{2}_{\{a,b\}}(z_1,z_2) 
+ f\left( \frac{z_1-1}{z_1 z_2} \right) K^{3}_{\{a,b\}}(z_1,z_2) \\
& + f \left( \frac{1- z_1 + z_1 z_2}{z_2} \right)
K^{4}_{\{a,b\}}(z_1,z_2) + f \left( z_1 (1- z_2) \right)
K^{5}_{\{a,b\}}(z_1,z_2) \, , 
\label{eq:TwoParticleOneZero} 
\end{aligned}    
\end{equation}
where
\begin{equation}
f(z) = g^4 \frac{(z+\bar{z}) - (\alpha -\bar{\alpha})}{2}  \,,  \label{eq:ffunctionstructure}
\end{equation}
and for $a=1$ and $b=0$
\begin{equation} 
\begin{aligned}
&K^{1}_{\{1,0\}}(z_1,z_2)  = F^{(2)}(z_1) \, ,  \quad  
 K^{2}_{\{1,0\}}(z_1,z_2) = (x^2_{il})^2 x^2_{kj} P_{m,il,jk}  \, , \quad \\
\quad & K^{3}_{\{1,0\}}(z_1,z_2) = x^2_{il} x^2_{im} x^2_{jk} L_{i,jk,lm} \, , \quad 
 K^{4}_{\{1,0\}}(z_1,z_2) = x^2_{il} x^2_{jk} x^2_{lm}
L_{l,jk,im} \, , \quad \\
& K^{5}_{\{1,0\}}(z_1,z_2) =F^{(2)}(z_1(1-z_2)) \, .  
\end{aligned}  
\end{equation} 
It is possible to obtain the result for $a=0$ and $b=1$
using the parity invariance of (\ref{eq:ParityInvariance}), but we are going to write it down explicitly for the readers convenience
\begin{equation} 
\begin{aligned}
&K^{1}_{\{0,1\}}(z_1,z_2)  = (x^2_{ik})^2 x^2_{ml} P_{j,ik,lm} \, ,  \quad  
 K^{2}_{\{0,1\}}(z_1,z_2) = F^{(2)}(z_2^{-1}) \, , 
 \\ 
& K^{3}_{\{0,1\}}(z_1,z_2) = F^{(2)}\left( \frac{z_1-1}{z_1 z_2} \right) \, , \quad  
 K^{4}_{\{0,1\}}(z_1,z_2) = x^2_{kj} x^2_{ik} x^2_{lm} L_{k,lm,ij} \, , \quad \\
& K^{5}_{\{0,1\}}(z_1,z_2) =x^2_{ij} x^2_{ik} x^2_{lm} L_{i,jk,lm} \, .  
\end{aligned}  
\end{equation} 
Notice that the structure for the two-loop 
two-particle contribution $\mathcal{M}^{(2)}_{2,\{a,b\}}(z_1,z_2)$ 
is similar to the structure of the one-loop two-particle result
given in (\ref{eq:Onelooptwoparticle}). In particular, 
both cases have five terms and the arguments 
of the functions $m$ are the same as the arguments 
of the functions $f$. This means that effectively
to go from one-loop to two-loop, one promotes the one-loop integrals to two-loop integrals keeping
the prefactors unchanged. This pattern seems to
be true to $l$ loops for any $l$ and for any values of $a$ and 
$b$ with 
\begin{equation}
a+b =l -1. 
\end{equation}
We have tested up to five loops that 
\begin{equation}
K^{1}_{\{a,0\}}(z_1,z_2)  = F^{(a+1)}(z_1) \, , \quad 
K^{5}_{\{a,0\}}(z_1,z_2) =F^{(a+1)}(z_1(1-z_2)) \, , 
\label{eq:allloop}
\end{equation} 
and we have also found $l$ loop integral representations 
for others $K^{i}_{\{a,b\}}$. We hope to report 
these and further results in a future publication. 

The other two-particle contribution involves 
only bridges of length zero and it is a highly 
constrained object. It is parity invariant 
(\ref{eq:ParityLengthZero}) and the sum of it with two
one-particle contributions is flip invariant or rotation invariant, see figure \ref{fig:TheFlippingFive}. 
Our result is given below using the following definition        
\begin{equation}
P_{00} \equiv \mathcal{M}^{(2)}_{1,\{ 0 \}}(z_1)
+ \mathcal{M}^{(2)}_{1,\{ 0 \} }(z_2)
+ \mathcal{M}^{(2)}_{2; \{0,0\}}(z_1,z_2) \, , 
\end{equation}
and 
\begin{align}
P_{00}&=h^{(2)}(z_1,z_2)+h^{(2)}\left(\frac{z_2}{1+z_1(z_2-1)},\frac{z_1-1}{z_1z_2}\right)+h^{(2)}\left(\frac{1}{z_2},z_1(1-z_2)\right)
\nonumber \\
&+h^{(2)}\left(\frac{z_1z_2}{z_1-1},\frac{1}{z_1}\right)+h^{(2)}\left(\frac{1}{z_1(1-z_2)},\frac{1+z_1(z_2-1)}{z_2}\right),
\label{TwoParticleZeroZero} 
\end{align}
with 
\begin{align}
h^{(2)}(z_1,z_2) &= \frac{1}{2} (\alpha_1(1-\alpha_2)-z_1(1-z_2)+\textrm{\bf{c.c.}})\left[ x^2_{ij} x^2_{jl} x^2_{lm} P_{k,lj,m,i}\label{functionm2}\right. \\
&\left.-x^2_{ij}x^2_{lm}(L_{j,mi,lk}x_{jk}^2+L_{l,mi,jk}x_{lk}^2)+2 F^{(2)}(z_1(1-z_2))\right] . \nonumber
\end{align} 

The two-loop result given above has the same structure as the one-loop result of (\ref{oneloopdecagon}), in other words, 
the two-results are given by a sum of one function 
evaluated at five different points. To go to one-loop
to two-loop one only has to correct this function.
It seems that (\ref{TwoParticleZeroZero}) is the general 
solution to flip invariance and at all loops one only
needs to correct the function $h$. We hope to
address this question in the future. 
There is one further property of all two-particle contributions: they vanish after twisting the polarizations  $\alpha_i\rightarrow z_i$. In fact this property together with the degree of the polarizations might prove that the same structure appear at all loops. In addition, all the results for the two-particle contributions are definite combinations
of Feynman integrals. This result is non-trivial 
because there was the possibility that Feynman integrals only show up when one computes a 
complete  correlation 
function and sums over all the mirror particles 
contributions.

\section{Planar correlation functions}\label{planarcorrelationfunctions}
The results described in the previous section will be used in the following to obtain correlation functions of half-BPS operators in $\mathcal{N}=4$ SYM. As we will see, the one and two particle contributions are enough to obtain a five point function with special polarizations and some four point functions.

\subsection{Five point function}
In the following, we will focus on planar five point correlation functions of large half-BPS operators  $L_i=2K\gg 1$.  The starting point of the hexagonalization procedure is the enumeration of all skeleton graphs which are obtained by tree level Wick contractions 
\begin{align}
&\langle \mathcal{O}_{2K}(x_i,y_i)\dots \mathcal{O}_{2K}(x_j,y_j) \rangle\bigg|_{\textrm{tree}} =\left(\frac{y_{im}^2y_{ml}^2y_{lk}^2y_{kj}^2y_{ji}^2}{x_{im}^2x_{ml}^2x_{lk}^2x_{kj}^2x_{ji}^2}\right)^K\sum_{l_{\nu\beta}} a_{l_{\nu\beta}}(N_c) G_{l_{il}l_{ik}l_{mk}l_{mj}l_{lj}}(z_i,\alpha_i) \, , \nonumber \\
&G_{l_{il}l_{ik}l_{mk}l_{mj}l_{lj}}(z_i,\alpha_i)= \left(\frac{u_2u_1\tau_1}{\sigma_1\sigma_2 v_1}\right)^{l_{km}}\left(\frac{\sigma_1 \tilde{t}}{u_1 t}\right)^{l_{jm}}\left(\frac{u_1}{\sigma_1}\right)^{l_{ik}}\left(\frac{\tau_2\sigma_1}{v_2u_1}\right)^{l_{jl}}\left(\frac{\sigma_2}{u_2}\right)^{l_{il}} \, , \label{eq:Gfunctiontreelevel}
\end{align}
where $a(N_c)$ are constants that depend on the number of colors and symmetry factors of each diagram and with the cross ratios\footnote{The space time cross ratios will not be used explicitly and so we do not give their definition. }
\begin{align}
\sigma_1=\frac{y_{i m}^2 y_{k l}^2}{y_{i k}^2 y_{l m}^2}, \, \tau_1=\frac{y_{i l}^2 y_{k m}^2}{y_{i k}^2 y_{l m}^2}, \, \sigma_2=\frac{y_{i l}^2 y_{j k}^2}{y_{i j}^2 y_{k l}^2}, \, \tau_2=\frac{y_{i k}^2 y_{j l}^2}{y_{i j}^2 y_{k l}^2}, \, t=\frac{y_{i k}^2 y_{j m}^2}{y_{i m}^2 y_{j k}^2} \, .
\end{align}

\begin{figure}[t]
\centering
\includegraphics[width=0.8\textwidth]{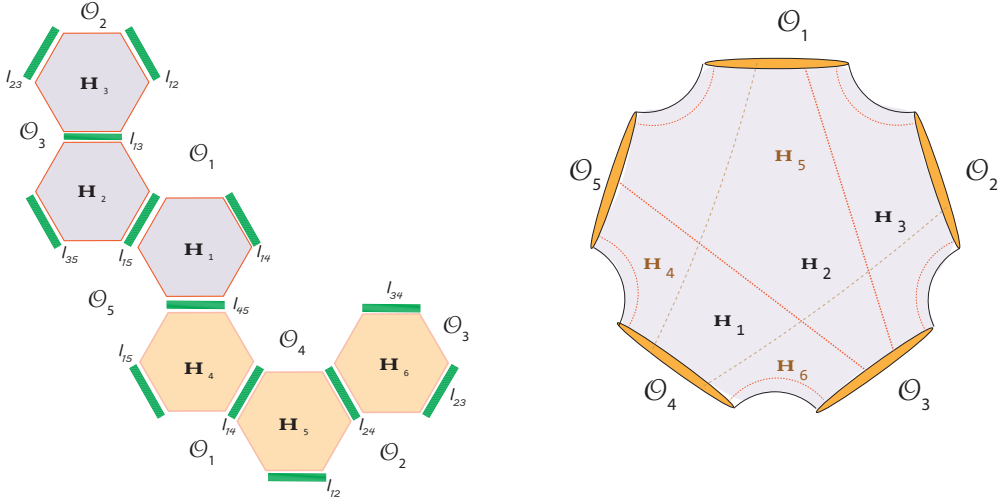}
\caption{Tessellation of a particular five point function skeleton graph where we assumed $l_{25}=0$. Notice that the configuration with all  $l_{ij}$ non-zero has a non-planar topology. After tessellating a given skeleton graph one is instructed to insert a complete set of mirror particle states to glue the hexagons back, see the figure on the left. }
\label{fig:tesselating2} 
\end{figure}

\begin{figure}[t]
\centering
\includegraphics[width=0.45\textwidth]{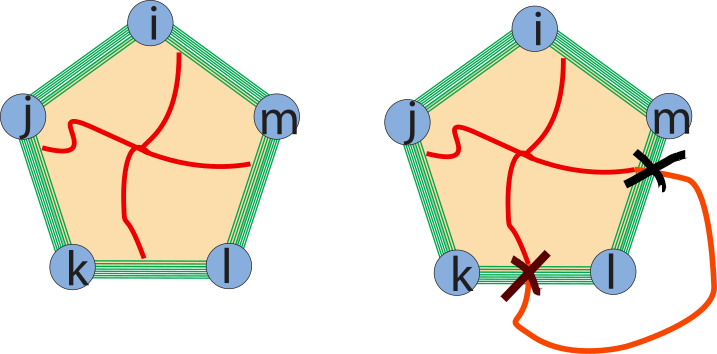}
\caption{At tree level the five point function is just given by (\ref{eq:Main5ptGraph}). As the coupling is turned on there are many different interactions that can take place. In principle, both figures on the left and right would be possible. However, since we take the operators to be large the figure on the right hand side is suppressed. Thus, we are left with interactions that live either on the inside or on the outside of the frame created by the five points. So the five point function is given by the square of the object we call the decagon which is nothing more than the contribution of these interactions on the inside of the frame.}
\label{fig:frame} 
\end{figure}

\begin{figure}[t]
\centering
\includegraphics[width=0.65\textwidth]{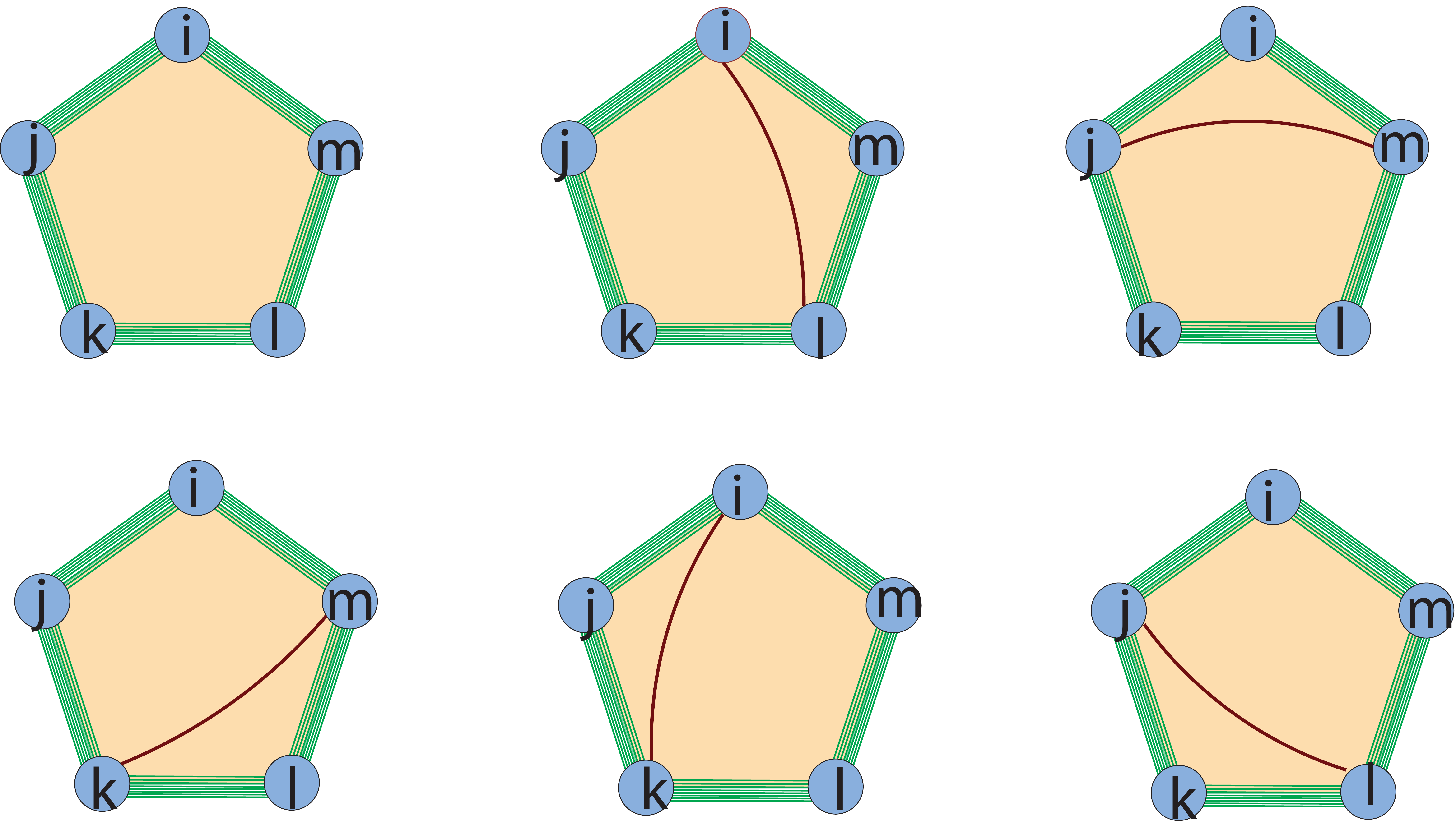}
\caption{Relevant (non-zero) neighbouring graphs to the decagon for both one and two particles. }
\label{fig:ParticleContributionsneighbouring} 
\end{figure}

 The next step is to tessellate each skeleton graph into six hexagons as shown in figure \ref{fig:tesselating2}. 
As pointed out in \cite{Frank1} the computation of a generic skeleton graph is complicated as it involves the contribution of multi-particle states in different edges. A simplification is obtained by projecting the five point function to a specific polarization in which case only the one-particle and two-particle contributions are relevant.  One such polarization is 
\begin{align}
y_i &= \{ 0,0,0,0,1,i \} \, , \quad 
y_m = \{ 0,0,1,i, \beta , -i \beta \}  \, , \quad
y_l = \{1, -i, 1, -i, 0,0 \} \, , \\
y_k & = \{ 1, i, 0,0,0,0 \} \, , \quad 
y_j = \{ 1, -i, 0,0, \gamma, - i \gamma \} \, .
\end{align}
As can be seen in figure \ref{fig:frame}  all corrections that connect the inside with the outside of the frame are suppressed in the limit $K\gg 1$. Consequently this specific five point function is given by the square of an object which will be called from now on, the decagon $\mathbb{D}$  
\begin{equation}
\frac{\mathbb{D}^2}{(x_{im}^2x_{ml}^2x_{lk}^2x_{kj}^2x_{ji}^2)^K}=\left( \frac{\partial}{\partial \beta} \right)^{K} 
\left( \frac{\partial}{\partial \gamma} \right)^{K} \langle \mathcal{O}_{2K} (x_i,y_i) 
\mathcal{O}_{2K} (x_m,y_m) \dots \mathcal{O}_{2K} (x_j,y_j)  \rangle \, \Big|_{\beta=\gamma=0} \label{eq:SimplestFivepointfunction}
\end{equation}      
Another way to see this is by looking into the tessellation of this five point function and to notice that these limits and polarizations  make the  the bridges $l_{ii+1}\approx K$\footnote{The $\approx$ shows up here because the effect of neighbouring graphs can mix different skeletons graphs contributing to a particular polarization as is discussed below.} which isolate the hexagons $1,2,3$ and $4,5,6$ in figure \ref{fig:tesselating2} from each other.  
At tree level there is just one graph that contributes to the simplest 
five point function of (\ref{eq:SimplestFivepointfunction}) which is 
\begin{align}
 d^K_{im} d^K_{ml} d^K_{lk} d^K_{kj} d^K_{ji}\label{eq:Main5ptGraph},
\end{align}
which should be dressed with the zero length one and two particle contributions for
loop corrections. 
The contribution of mirror particle states  have $R$-charge which can change the polarization of a given skeleton graph to (\ref{eq:Main5ptGraph}). As we have seen in the previous section  the hexagonalization of any of the graphs  in \ref{fig:ParticleContributionsneighbouring} apart from the first can be written as a combination of seven polarization structures 
\begin{align}
&f\left(\frac{1}{z_1}\right)\mathcal{P}_{\frac{1}{\alpha_1}}+f(z_2)\mathcal{P}_{\alpha_2}-f\left(\frac{z_1-1}{z_1z_2}\right)\mathcal{P}_{\frac{1}{\alpha_1\alpha_2}}-f\left(\frac{1-z_1+z_1z_2}{z_2}\right)\mathcal{P}_{\frac{\alpha_1}{\alpha_2}}-f(z_1(1-z_2))\mathcal{P}_{\alpha_1\alpha_2} \nonumber\\
&+f(z_1)(\mathcal{P}_{\alpha_1}+\mathcal{P}_{\frac{\alpha_1}{\alpha_2}}+P_{\alpha_1\alpha_2})+f\left(\frac{1}{z_2}\right)(\mathcal{P}_{\frac{1}{\alpha_2}}+\mathcal{P}_{\frac{1}{\alpha_1\alpha_2}}+\mathcal{P}_{\frac{\alpha_1}{\alpha_2}}).
\label{eq:Structures}\
\end{align}
where $\mathcal{P}_x$ represents the one or two particle contribution with one or zero length bridges and the function $f(z)$ was defined in  (\ref{eq:ffunctionstructure}). 
The structures in the first line transform among each other upon a cyclic rotation of the decagon, {\it{ i.e.} $i\rightarrow m,\,m\rightarrow l\dots$ } while the two on the second line transform differently under the cyclic rotation\footnote{It is then easy to understand why the combination multiplying both $f(z_1)$ and $f(\frac{1}{z_2})$ vanishes in the tessellation of a decagon with zero length bridges.}. 
The relevant skeleton graphs that contribute to the correlation function (\ref{eq:SimplestFivepointfunction}) are given by the solutions of (see figure \ref{fig:ParticleContributionsneighbouring} for all possible nonzero graphs\footnote{Both one and two particle contribution can only change the R-charge structure by two units, so it is enough to consider only the graphs in \ref{fig:ParticleContributionsneighbouring}.})
\begin{align}
G_{l_{il}l_{ik}l_{mk}l_{mj}l_{lj}}(z_i,\alpha_i)\times f(z_i)=1 \, , \label{eq:equationforneighbouring}
\end{align}
where $f$ is one of the seven structures that appear in the tessellation of the skeleton graphs\footnote{The $1$ on the right hand side of (\ref{eq:equationforneighbouring}) appears because we have normalized (\ref{eq:Gfunctiontreelevel}) with the only prefactor that is relevant for the polarization that we want. }, $G$ is defined in (\ref{eq:Gfunctiontreelevel}). As an example, pick the second term in (\ref{eq:Structures}) which can be written as $1+\sigma_2-\tau_2$ and consider it on (\ref{eq:equationforneighbouring}). It is straightforward to see there is no solution to this equation. Since the first five terms in (\ref{eq:Structures}) transform into each other under rotation there is no solution of (\ref{eq:equationforneighbouring}) for any of them. A similar and small computation implies that there are two other solutions $l_{il}=1,l_{\nu\beta\neq il}=0$ and $l_{ik}=1,l_{\nu\beta\neq ik}=0$ for the remaining structures in (\ref{eq:Structures}). 

For example the second graph in the second line in figure \ref{fig:ParticleContributionsneighbouring} where $l_{ik}=1$ the only nontrivial component comes from $f(z_1)$ and contributes to the correlator (\ref{eq:SimplestFivepointfunction}) as
\begin{align}
z_1\bar{z}_1 (\mathcal{P}_{\alpha_1}+\mathcal{P}_{\frac{\alpha_1}{\alpha_2}}+\mathcal{P}_{\alpha_1\alpha_2})=z_1\bar{z}_1P\left(\frac{z_2-1}{z_2},\frac{1-z_1(1-z_2)}{z_1z_2}\right).
\end{align}
The contribution from the other graphs can be obtained by a cyclic rotation. It is now simple to obtain the final result for the correlator 
\begin{align}
\mathbb{D}&=g^2 \big[(z_2+\bar{z}_2-1)\frac{F^{(1)}(z_2)}{2}-z_1\bar{z}_1F^{(1)}(z_1)\big]\nonumber\\
+&g^4 \big[\frac{(1-z_2-\bar{z}_2)}{2}\left(\frac{P(1-z_1(1-z_2)),1-z_1)}{z_2\bar{z}_2}-\frac{z_1\bar{z}_1L(\frac{1}{1-z_2},z_1)+L(1-z_2,\frac{1}{z_1})}{(1-z_1(1-z_2))(1-\bar{z}_1(1-\bar{z}_2))}\right.\nonumber\\
+&2F^{(2)}(z_2)\bigg)+\frac{z_1\bar{z}_1}{2}P\left(\frac{z_2-1}{z_2},\frac{1-z_1(1-z_2)}{z_1z_2}\right)+z_1\bar{z}_1F^{(2)}(z_1)\big]+\textrm{cyclic rotations}\label{eq:twoloopresult}.
\end{align}

	In \cite{DrukkerPlefka}, it was shown how to obtain any one loop $n$-point function of half BPS operators. We have checked that the integrability method laid out above reproduces exactly the one from \cite{DrukkerPlefka} at one-loop. The two loop correction for our five point function was not known in the literature. However we manage to compute it with a different method that uses both Lagrangian insertion method \cite{EdensII} and chiral algebra twist\cite{ChiralSymmetry,Goncalves:2019znr}. The details will appear elsewhere\cite{LagrangianInsertionMethod}. 

%
%
%
%
%


\subsection{Four-point 
functions}  
\begin{figure}[t]
\centering
\includegraphics[width=0.4\textwidth]{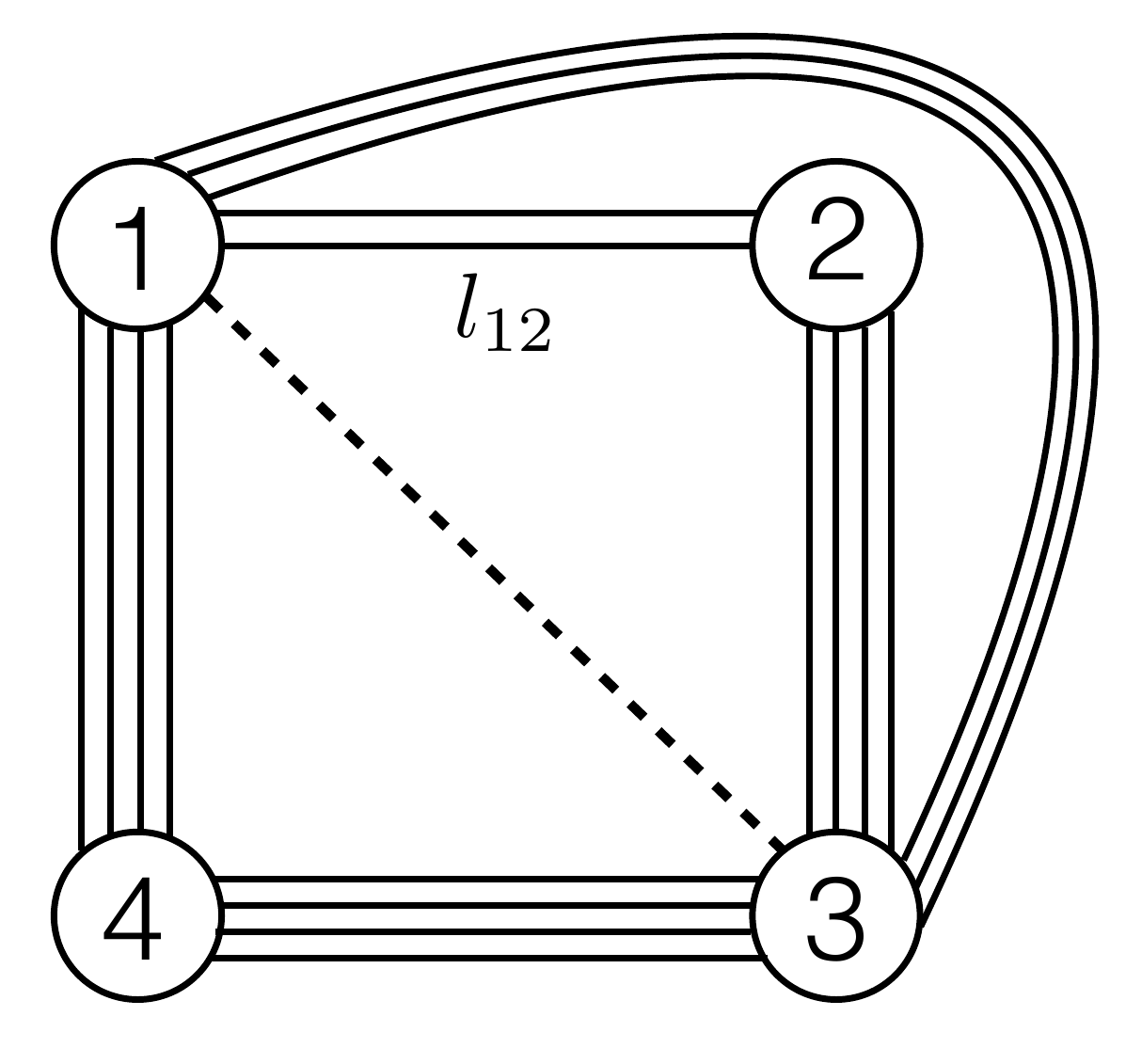}
\caption{
An example of a tree-level contribution to the four-point functions computed 
using integrability. 
The other tree-level graphs are obtained by 
moving some of the lines connecting the 
operators $\mathcal{O}_1$ and $\mathcal{O}_3$ 
to the dashed line and there are $K$ of them. 
For $l_{12}=0$ there are $K-1$ additional graphs 
obtained by permuting $\mathcal{O}_1$ and $\mathcal{O}_4$ in the figure 
and moving the propagators.      
The bridge length $l_{12}$  
will varry from case to case and the polarizations 
are such that $y_{24}=0$, see (\ref{eq:Polarizations}). As explained
in the main text, using the two-particle 
integrability result computed in this paper
we can evaluate $l_{12}<3$ up to two-loop and 
$l_{12}=l$ 
up to $l$ loops. Notice that it is important to have non-zero  connections between the  
operators $\mathcal{O}_1$ and $\mathcal{O}_3$ 
in order to suppress two new integrability contributions: three-particle in different edges 
and the loop over an operator which correspond to a four-particle contribution. 
}
\label{fig:FourPointsFigure} 
\end{figure}

The knowledge of both one-particle and two-particle contributions enables one to compute 
several four-point functions as well. It is easy to reduce the two-particle contribution from five operators
to four  by identifying a pair of operators. In all the needed cases the reduction 
of the integrals does not produce any divergence and it can be done easily by also
identifying points. 
In this section, we are going to compute a few
planar four-point functions at two- and three-loops of half-BPS operators 
using integrability and we will compare the results
with the perturbative ones obtained in 
\cite{3loopdata}. Notice that the correlators considered in this section are known up to five-loops \cite{45loopdata}. The comparison is a strong test of our integrability computations of section 
\ref{sec:Integrability}.  
Specifically, we are going to consider the following class of operators (the setup is shown in figure \ref{fig:FourPointsFigure}) 
\begin{equation}
\begin{aligned}
\mathcal{O}_1 &= {\rm{Tr}} (Z^{K} X^{l_{12}} Y^{K}) + {\rm{perm}} \, , \quad   \quad \mathcal{O}_2 = {\rm{Tr}} (\bar{X}^{K+l_{12}}) \, ,  \\
 \mathcal{O}_3 &= {\rm{Tr}} (Z^{K} X^{K} \bar{Y}^{K} )+ {\rm{perm}}  \, , \quad \quad \, \,
\mathcal{O}_4 = {\rm{Tr}} (\bar{Z}^{2K}) \, ,   
\label{eq:OperatorsFour} 
\end{aligned}
\end{equation}
with $K \gg 1$ and $l_{12}=0,1,2,3$. The cases $l_{12}=0,1,2$ are going to be computed
at two-loop and the remaning case $l_{12}=3$  at three-loop\footnote{The minimum value of
$K$ depends on the loop order. To suppress unwanted multi-particle contributions, $K$ has to
be greater than $2$ for the two-loop cases ($l_{12} < 3$) and greater than $3$ for the three-loop case ($l_{12}=3$).}. In fact, we can compute  
the case $l_{12}=l$ 
up to $l$ loops as it needs only the one-particle contributions and one component 
of the two-particle at $l$ loops. The needed component is precisely the one 
determined at (\ref{eq:allloop}), see the discussion below.  

Note that the polarizations and the lengths 
of the operators in (\ref{eq:OperatorsFour}) were judicious chosen in order for 
the integrability computation to only involves the one-particle and two-particle contributions. 
It is important that the operators $\mathcal{O}_1$ and $\mathcal{O}_3$ are 
connected at tree-level (they have the $Y$  and the $\bar{Y}$ fields respectively) otherwise 
there will be additional integrability contributions.
For example, consider the four-point function of four half-BPS operators of length two (the 
so called $20^{\prime}$ operators). Using the rule described in figure \ref{fig:ThePowerofg} for determining at each order in
$g$ a multi-particle contribution kicks in, one can see that the computation of the $20^{\prime}$'s correlation function needs one-, two-, three- and four-particle contributions at two-loop
(at one loop it only needs one- and two-particle contributions).  
Note that the four-particle contribution closes to form a loop. 
The three-particle contribution is only known
at one-loop (in fact any string of mirror particles
is known by recursion relations at one-loop, see \cite{HandlingII}) and 
its knowledge at two-loop would enable one 
to also compute the dodecagon or a special polarized six-point function.  
On the other hand, almost noting is know about the mirror loops apart 
from the fact that its leading contribution is 
zero because of supersymmetry (one is wrapping a BPS operator).  

The four-point function of the operators  
(\ref{eq:OperatorsFour}) are obtained 
by setting the polarizations vectors as    
\begin{equation}
\begin{aligned}
y_1 &= \{ 1, i, \alpha, i \alpha , \beta, i \beta\}   \, , \quad \quad & y_2 = \{ 0,0,1,-i,0,0\} \, ,  \\
y_4 &= \{1, -i,0,0,0,0 \}  \, , \quad \quad  
&y_3 = \{ 1,i, 1,i, 1 , -i  \}  \, ,  
\label{eq:Polarizations}
\end{aligned}
\end{equation}
and applying the following differential operators in the correlation functions
\begin{equation}
G_{1234}  = \left( \frac{\partial}{\partial \beta} \right)^{K} \left( \frac{\partial}{\partial \alpha} \right)^{l_{12}} 
\langle \mathcal{O}_{1} (x_1) \mathcal{O}_{2} (x_2) \mathcal{O}_{3} (x_3) 
\mathcal{O}_{4} (x_4) \rangle \, \Big|_{\alpha=\beta=0} \, .  
\end{equation}         
The perturbative result for the correlators  
$G_{1234} = \prod_{i=1}^4 \sqrt{L_i}  \, G^{\prime}_{1234}/N_c^2$
at the necessary loop order 
can be read from \cite{3loopdata}. The expressions depend only on 
the ladders integrals $F^{(L)}(z, \bar{z})$ of (\ref{eq:LadderIntegralResult})  and they are 
given in our conventions by        
\begin{equation}
\begin{aligned}
&G^{\prime}_{1234} \Big|_{l_{12}=0}^{g^4}   
= (1-z)(1-\bar{z}) F^{(2)}[1-z] + F^{(2)}\left[\frac{z}{z-1} \right] \, ,  \\
&G^{\prime}_{1234} \Big|_{l_{12}=1}^{g^4}  = 
(z + \bar{z} - 2 z \bar{z}) F^{(2)}[1-z] + \frac{(1- z\bar{z})}{(1-z)(1-\bar{z})} F^{(2)}\left[ \frac{z}{z-1} \right] \, ,  \\
&G^{\prime}_{1234} \Big|_{l_{12}=2}^{g^4}  = 
z \bar{z} \,  F^{(2)}[1-z] + \frac{F^{(2)}\left[ \frac{z-1}{z} \right]}{z \bar{z}} 
\, ,  \\
&G^{\prime}_{1234} \Big|_{l_{12}=3}^{g^6}  = 
z  \bar{z} \, F^{(3)}[1-z] - 3  \frac{ F^{(3)}\left[\frac{z-1}{z}\right] }{z \bar{z}} \, ,  
\label{eq:perturbativedata} 
\end{aligned} 
\end{equation}
where the cross-ratios $z$ and $\bar{z}$ were 
defined in (\ref{eq:Crosszzb}).  
\begin{figure}[t]
\centering
\includegraphics[width=0.7\textwidth]{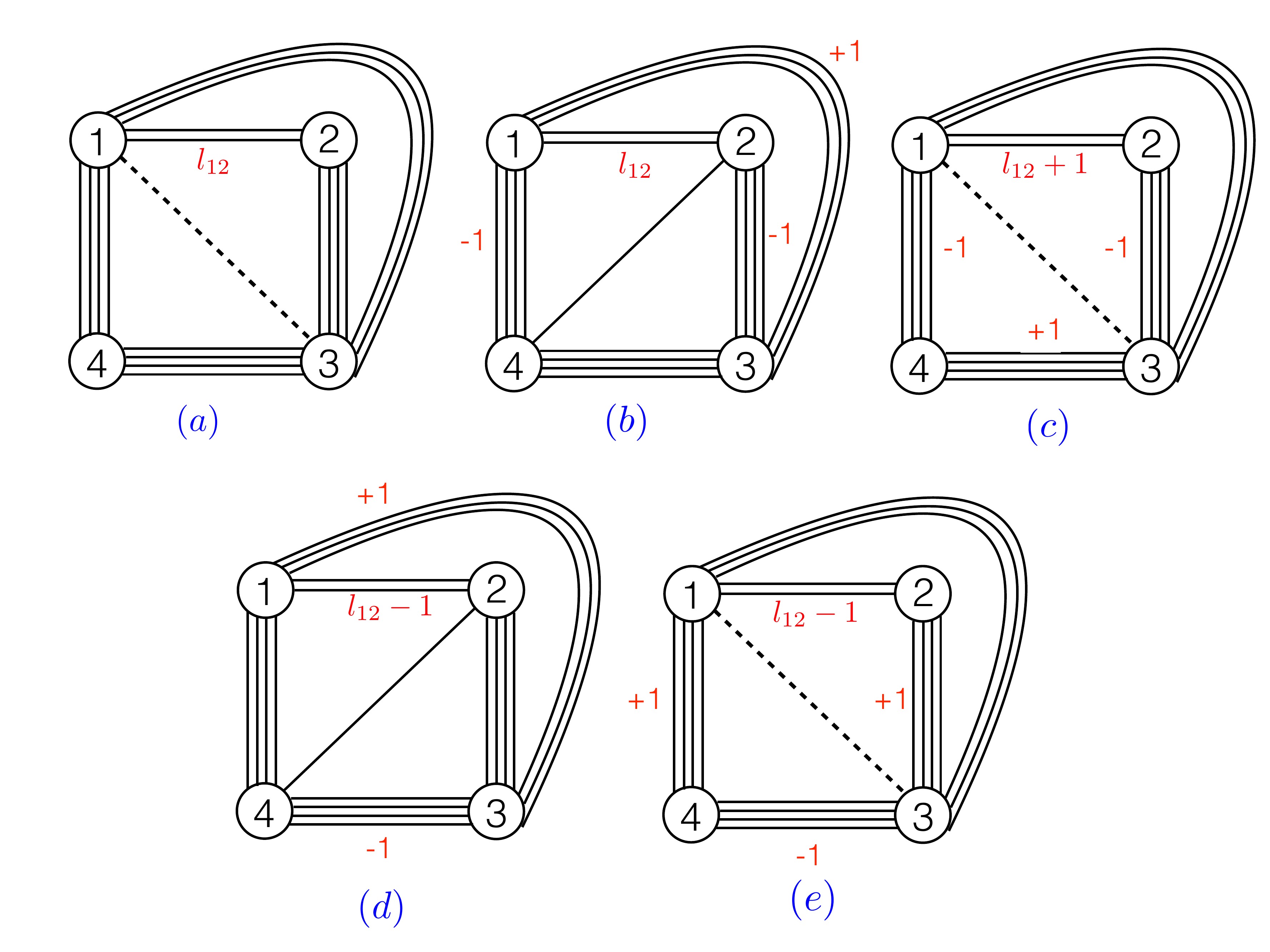}
\caption{
All the propagators structures used for computing 
the four-point functions using integrability. 
The graph $(a)$ has the tree-level propagator structure and all the others are 
neighboring graphs. 
To find all the graphs from the diagrams above 
one has to connect the four operators in all inequivalent ways 
keeping both planarity and the propagator structure showed. 
For example, four graphs of case $(a)$ give a nontrivial contribution for   
$l_{12} \neq 0$ at two-loop (at three-loop or $l_{12}=3$ the number of graphs is six 
and for $l_{12}=0$ there are additional graphs by permuting the operators 
$\mathcal{O}_1$ and $\mathcal{O}_4$).
The four graphs are obtained by 
moving the propagators connecting the operators 
$\mathcal{O}_1$ and $\mathcal{O}_3$, they are:
 the graph showed, the graph when there is one propagator 
 inside the middle square, the graph when all the propagators are inside the middle square and the graph when only one propagator is outside the square. 
Similarly, the cases $(b),(c),(d)$ and $(e)$ 
give rise to several graphs.
The contribution of each graph is computed by using the reduced to four-points two-loop two-particle contribution of this paper.     
}
\label{fig:Vizinhos4} 
\end{figure}

To compute the correlators above using integrability, we follow the same steps as in the 
five-point function computation. Besides evaluating the mirror particle corrections 
to the tree-level graphs, it is also necessary to
deform a bit the polarizations and compute some
neighboring graphs. 
The neighboring graphs give a finite 
contribution after sending the deformation to zero at the very end because
the mirror particles contributions carry $R$-charge and consequently they change the $R$-charge structure
of the original graphs. 
To perform the computations one 
reduces the two-loop two-particle contributions 
given in (\ref{eq:TwoParticleOneZero}) and (\ref{TwoParticleZeroZero}) to the case of 
four-points by identifying the fifth point with one
of the four-points. The precise identification depends on the graph being computed. 
As mentioned before, the integrals can be easily reduced as no divergence 
shows up and one can identify 
the points on the level of the integrand and then
perform the integral after the identification which always give in our cases a ladder two 
computed for a specific cross-ratio. 
If one expresses the 
four-point $R$-charge cross-ratios as 
\begin{equation}
\alpha \bar{\alpha} = \frac{X}{Y}  \, z \bar{z}   \, , \quad  \quad (1-\alpha)(1-\bar{\alpha}) = \frac{Z}{Y} \, (1-z)(1-\bar{z})  \, , 
\label{eq:Crosszzb} 
\end{equation} 
with 
\begin{equation}
X = \frac{y_{12} y_{34}}{x_{12}^2 x_{34}^2} \, , \quad 
Y = \frac{y_{13} y_{24}}{x_{13}^2 x_{24}^2} \, , \quad 
Z = \frac{y_{14} y_{23}}{x_{14}^2 x_{23}^2} \, ,
\end{equation}
one verifies that all reduced one-particle and two-particle contributions are first order polynomials   
in $A/B$ where both $A$ and $B$ are $X,Y$ or $Z$. This 
implies that the mirror particles can only change the $R$-charge structure of a graph by two units.
Thus it is only possible to have neighbouring graphs 
which differ from the original graph by four propagators and each line can differ from the tree-level graphs by at most one propagator. The relevant graphs, more precisely the propagator structures, are shown in figure
\ref{fig:Vizinhos4}. As two examples, let us explicitly compute the graphs $(a)$ and 
$(c)$ of the figure \ref{fig:Vizinhos4} for $l_{12}=1$.  Dividing each graph by the unique 
tree-level propagator structure, we have the following contributions to the final result using the expressions and notations of 
section \ref{sec:Integrability}
\begin{equation}
\begin{aligned}
(a): \;  & \mathcal{M}^{(2)}_{1,\{ 0 \}}\left(\frac{z}{z-1}\right)
+ \mathcal{M}^{(2)}_{1,\{ 1 \} }(1)
+ \mathcal{M}^{(2)}_{2; \{1,0\}} 
\left(1, \frac{z}{z-1} \right) \Bigg|_{\, {\rm{zero}} \, R-{\rm{charge}}} \\
& = g^4 (1+z+ \bar{z}-2 z \bar{z}) \left(  \frac{1}{2} F^{(2)}[1-z] + \frac{F^{(2)}\left[ \frac{z-1}{z} \right]}{z \bar{z}} \right) \, . 
\end{aligned} 
\end{equation}
where we have used that $\mathcal{M}^{(2)}_{1,\{ 1 \} }(1)=0$ and 
\begin{equation}
\begin{aligned}
(c): \;  & \mathcal{M}^{(2)}_{1,\{ 0 \}}\left(\frac{z}{z-1} \right) \Bigg|_{\, {\rm{component}} \, Z/X} \\
& = \, g^4 F^{(2)}\left[ \frac{z}{z-1} \right] \, . 
\end{aligned} 
\end{equation}
Continuing in this way and summing all the integrability contributions one gets agrement with the perturbative results given in 
(\ref{eq:perturbativedata}). 

Note that the calculation at three-loop for $l_{12}=3$ was possible because there is 
only two graphs which have two-particle corrections for this case. They are inside case $(e)$ of
figure \ref{fig:Vizinhos4}. Precisely one needs 
$\mathcal{M}^{(3)}_{2; \{2,0\}}$ and $\mathcal{M}^{(3)}_{2; \{0,2\}}$.   These
two-particle contributions have the structure given in (\ref{eq:TwoParticleOneZero})
and due to the $R$-charge  we only need to know for the computations of the four-point correlator the functions $K^5_{\{2,0\}}$ and
 $K^3_{\{0,2\}}$ which were determined in (\ref{eq:allloop})\footnote{The function 
 $K^3_{\{0,a\}}$ can be obtained from $K^5_{\{a,0\}}$ because of the parity invariance of
(\ref{eq:ParityInvariance}).}. Similarly, it is possible to compute the case $l_{12}=l$ up to $l$ loops only using (\ref{eq:allloop}) and the expressions for multi-particle contributions
in the same edge of \cite{Frank1}. 
We are not going to write the explicitly result here, because we do not know 
a closed expression for any $l$ and the computation is straightforward. 
For $l$ greater than five, the result is new.

\section{Conclusion \label{sec:IntegrabilityCalculation}}

In this paper, we have computed the simplest five point function in $\mathcal{N}=4$ SYM at two-loops using integrability. 
The result is expressed in terms of ladders and pentaladder integrals which were computed for the first time in this paper.  
To obtain this result we have computed all 
the two-loop two particle integrability contributions and derived some $l$ loop results along the way. 
The knowledge of the two particle contribution enables us also to compute a set of  
four point functions by reducing the results from
five to four points. The reduction is straightforward since
no diverge appears. 

As mentioned in the introduction, one of the goals of our two-loop computation is also to take the first steps  
towards a bootstrap approach to the decagon.  
One important question is what is the all loop 
basis of Feynman integrals. 
There are obvious generalizations of both the ladder and  pentaladder with five points and the differential equations derived in the Appendix \ref{Feynmanintegrals} can be used to study these integrals
but it is not clear if this is enough  
to cover the space of functions of this special five point function. 
It would be interesting to check this at higher loops. 
Recall that starting at three loops new kinds of contributions appear. 
One new contribution is the three-particle contribution with two particles in the same edge and the third one in an adjacent edge, see table \ref{table} for more cases.
This kind of contribution has never been analyzed  and it would be interesting to obtain the integrand for this case and check which integrals it generates. 
In this work, we have computed the decagon with all the five points on a given
space-time and $R$-charge plane. In this restricted kinematics one of the cross-ratios
is expressed as a function of the others. 
It is possible to lift this restriction and explore the full kinematical space. 
The only modification needed in the integrability calculation is to include 
in the weight factor generators that move points outside the plane. 
In principle there is no new integral appearing and one could reproduce the 
out of plane double box result of \cite{MatthiasPrivate}. One motivation for studying out of 
plane configurations is that the differential equations obeyed by the integrals
simplifies and it is possible to study them at the level of the integrability integrand. 

The multi-particle integrability contributions are very constrained objects. 
For example, the knowledge of just one $R$-charge structure of 
the two-particle length zero enables one to fix this contribution entirely by using
the flip invariance. We believe that exploring and solving all the symmetries will be important for higher loop integrability computations. 
There are more constraints that one can try to explore. 
It is known that any $n$-point function of half-BPS operators does not receive 
any quantum corrections if one applies the  Drukker-Plefka twist \cite{DrukkerPlefka,Superprotect}. This constraint is trivially implemented in the
hexagonalization approach because when imposing the twist the functions $f(z)$ of 
(\ref{eq:TwoParticleOneZero})
and all the prefactors of the functions $h^{(2)}(z_1,z_2)$  given in (\ref{functionm2}) 
vanishes identically. This property can be traced back to the form of
the weight factor. However, there is an additional twist that seems to be realized 
non trivially, the chiral algebra twist of \cite{ChiralSymmetry}.
It would be interesting to check what kind of constraints this puts on the different 
multi-mirror particle contributions. 

In this work, we have computed the mirror particle contributions by fitting 
series expansions on the cross-ratios by a basis of integrals. It will 
be great to develop a deeper understanding of the integrability integrand.
An interesting analysis of the one loop integrand has been put forward in \cite{deLeeuw:2019qvz,deLeeuw:2019tdd}. The one loop result has been known for a while however the approach taken in that papers might shed some light on the space of functions  that the integrand integrates to. It is possible that the integrand can be great simplified 
and some of its summation done explicitly. In this case, maybe it would be possible
to evaluate it numerically. The quantum spectral curve and the SoV basis has 
been applied to the computation of correlations functions recently
\cite{QuantumSpectral,QuantumSpectralII,QuantumSpectralIII,DerkachovI,DerkachovII,BelitskySOV,GiombiShota}. It is likely that these techniques could help in the simplification of 
the integrand.

\section*{Acknowledgement}
We would especially like to thank Shota Komatsu for collaboration during the initial stages of this work and for discussion. 
We acknowledge useful discussions with Till Bargheer,  João Caetano, 
Burkhard Eden, Alessandro Geourgoudis, Matt von Hippel, Erik Panzer, 
Emery Sokatchev, Pedro Vieira and Matthias Wilhelm. 
We thank Frank Coronado and Shota Komatsu for comments on the draft. 
TF would like to thank the warm hospitality of the King's College London where part
of this work was done.
This work was supported by
the Serrapilheira Institute (grant number Serra-1812-26900).
The work of V.G. was supported by FAPESP grant 2015/14796- 7 and by the Coordenacao de Aperfeicoamento de Pessoal de Nivel Superior - Brasil (CAPES) - Finance Code 001.



\appendix


 \section{Finite coupling expressions and the one-particle contribution} \label{weakexpansion}
The goal of this section is to collect the finite coupling expressions involved in the integrability integrand (\ref{eq:Formula2twoparticle}). 
They can be easily expanded to compute the integrand up to any desired
loop order. The mirror bound-state $S$-matrix will not be given here and it 
can be derived by following the procedures described in \cite{HexagonalizationII}. 
Most of the expressions are given in terms of Zhukowsky variables $x$ defined by 
the relation
 \begin{equation}
 \frac{u}{g} = x + \frac{1}{x} ,
 \end{equation}
 and one selects the solution with good weak coupling behavior. 
The energy and momenta of a mirror particle bound state with length
$a$ is given by 
\begin{align}
 e^{- \tilde{E}_a} =
 \frac{1}{x^{[a]} x^{[-a]}} \; , \quad  \quad    \quad
 \tilde{p}_a =  - i \left[ \frac{a}{2} + \frac{g}{i} \left(\frac{1}{x^{[-a]}} - x^{[a]} \right) \right] \, , 
 \end{align} 
 where we used the short hand notation
 \begin{equation}
 f^{[a]} = f\left(u + i \frac{a}{2} \right). 
\end{equation}
In addition, we have for a physical magnon
\begin{align}
e^{i p_a} = \frac{x^{[a]}}{x^{[-a]}} \; , \quad \quad \quad 
E_{a} = \frac{1}{2} + \frac{g}{i} \left(\frac{1}{x^{[-a]}}-\frac{1}{x^{[a]}} \right) \, . 
\end{align} 
The measure is given by 
\begin{align} 
 \mu_a(u^{\gamma}) =  \frac{a (x^{[a]} x^{[-a]})}{g^2
 ( x^{[a]} x^{[-a]} -1)^2 ((x^{[a]})^2-1) ((x^{[-a]})^2-1)} \, , 
\end{align} 
and the dynamical part of the hexagon form factor for the case of 
fundamental particles  
can be written in a compact form as 
\begin{equation}
 h(u,v) = \frac{x_u^- -x_v^{-}}{x_u^-  - x_v^+}
 \frac{1-1/x_u^- x_v^+}{1- 1/x_u^+ x_v^+} \frac{1}{\sigma_{u v}} 
\end{equation}
with $\sigma_{u v}$  the BES dressing phase\cite{Beisert:2006ez}. The fused  transitions relevant for this paper are obtained as 
 \begin{equation}
 h_{ab} (u,v) = \prod_{k= - \frac{a-1}{2}}^{\frac{a-1}{2}}
 \prod_{l = - \frac{b-1}{2}}^{\frac{b-1}{2}} h(u^{[2k]}, v^{[2l]}) \, . 
 \end{equation}
The dressing phase in the mirror-mirror kinematics was derived in \cite{Arutyunov:2009kf} and is given by
\begin{align}
\sigma_{ab}(u^{\gamma},v^{\gamma}) &= \Phi(x^{+},y^{+})+ \Phi(x^{+},y^{-})+\Phi(x^{-},y^{+})+\Phi(x^{-},y^{-})\nonumber\\
&+\psi(x^{+},y^{+})+\psi(x^{-},y^{-})+b(x^{+},y^{+})+c(x^-,y^-)\\
&+\ln\frac{\Gamma(1-\frac{a}{2}+iu)\Gamma(1+\frac{a-b}{2}-i(u-v))\Gamma(1+\frac{b}{2}-iv)}{\Gamma(1+\frac{a}{2}-iu)\Gamma(1-\frac{a-b}{2}+i(u-v))\Gamma(1-\frac{b}{2}-iv)} \, , 
\end{align}
\begin{align}
&\Phi(x,y)=\sum_{n,m} \int \frac{dt}{t(e^t-1)}J_{1+m}(gt)J_{1+n}(gt)\frac{2(-1)^m\sin \pi\frac{m+n}{2}}{x^{1+n}y^{1+m}}+\int\frac{e^{-t}g\,dt}{2t}\bigg[\frac{1}{y}-\frac{1}{x}\bigg]\, , \\
&\Psi(x,y)= 2(-1)^{m+1}\sum_{m}\int \frac{dt}{t(e^t-1)}J_{m+1}(gt)\left(\frac{\cos(\frac{1}{2}(m\pi +(\frac{ia+u}{2})))}{y^{m+1}}-\frac{\cos(\frac{1}{2}(m\pi +(\frac{ib+v}{2})))}{x^{m+1}}\right) \, ,\nonumber\\
&b(x,y)=\int\frac{e^{-t}g\,dt}{t}\bigg[\frac{2}{x}+x-\frac{2}{y}-y\bigg],\, \ \ \ c(x,y)=\int\frac{e^{-t}g\,dt}{t}\bigg[\frac{1}{x}-\frac{1}{y}\bigg] \, , \nonumber
\end{align}
where each sum runs from $\frac{\pm-|\pm|}{2}$ to $\infty$. 

We also provide the weak coupling expansion of the integral of the one particle contribution at any loop order \cite{Frank1}
\begin{align}
\mathcal{M}_{1,\{l\}}(z)=\sum_{i=1}^{\infty}\frac{(-4)^{i-1}\left(\frac{1}{2}\right)_{i-1}}{\left(1\right)_{i-1}}\frac{(1-i)_{l}}{(i)_{l}} \, (m^{(i)}(z)+ m^{(i)}(z^{-1})) \, , 
\label{oneparticlegeneralformula}
\end{align}
where $l$ is the bridge length.

\section{Feynman integrals}\label{Feynmanintegrals}
The integrability integrand  associated with more than one particle on different mirror edges is a complicated object that depends in an intricate way  on the bound state number, rapidities and cross ratios.  Its contribution enters to the correlation function after summing and integrating over the bound states indices and rapidities respectively. However, the best one can do for now is to truncate the sum over the bound states indices up to some number and then do the integration over the rapidities.  This is the same as computing the power series expansion in terms of cross ratios up to some order. To have a better control over the full function of cross ratios we have matched the power series expansion against  a set of Feynman integrals for which we have explicit expressions in terms of Goncharov functions. 

We have focused, in this paper, on the two loop contribution to the two particle integrand in a special plane kinematics which depends on two complex cross ratios. It is then natural to consider a set of two loop conformal integrals in position space depending on five external points with a two dimensional kinematics. The two simplest finite integrals involving five points are 
\begin{align}
&L_{1,24,35}=\int \frac{d^4x_6d^4x_7}{x_{16}^2x_{27}^2x_{36}^2x_{67}^2x_{47}^2x_{56}^2x_{17}^2}=\frac{L(z,\bar{z},w,\bar{w},h)}{x_{12}^2x_{15}^2x_{34}^2} \, , \label{eq:fivepointLadderintegral}\\
&P_{1,34,25}=\int\frac{d^4x_6d^4x_{7}\,x_{17}^2}{(x_{16}^2x_{46}^2x_{36}^2)x_{67}^2(x_{47}^2x_{37}^2x_{27}^2x_{57}^2)}=\frac{P(z,\bar{z},w,\bar{w},h)}{(x_{34}^2)^2x_{25}^2} \, ,
\end{align}
where $h$ is associated with the distance out of the plane
\begin{align}
&z\bar{z}=\frac{x_{12}^2x_{34}^2}{x_{13}^2x_{24}^2} \, ,\,\,(1-z)(1-\bar{z})=\frac{x_{14}^2x_{23}^2}{x_{13}^2x_{24}^2}\, \,\,w\bar{w}=\frac{x_{15}^2x_{34}^2}{x_{13}^2x_{45}^2}\,, \, \,(1-w)(1-\bar{w})=\frac{x_{14}^2x_{35}^2}{x_{13}^2x_{45}^2}\, ,\nonumber \\
&\frac{x_{25}^2x_{14}^2}{x_{12}^2x_{45}^2}=\frac{w\bar{w}-\bar{w}z-w\bar{z}+z\bar{z}+h}{z\bar{z}}\nonumber.
\end{align}
The first integral can be thought as a  simple generalization of the ladder integral with four points (in fact one can get back to it  by taking $x_5$ equal to $x_2$ or $x_4$). To our knowledge these two integrals have never been computed in the literature\footnote{Let us remark that these integrals have been computed in other kinematical regimes where the number of cross ratios is less than four.} and so in the rest of this section we will sketch the main steps involved in their computation. 

Fortunately each of these two integrals in a plane can be computed in parametric space by direct integration of the Schwinger parameters. For example
\begin{align}
\lim_{x_4\rightarrow \infty}x_4^2 L_{1,24,35}^{(2)}=\prod_{i=1}^{6}\int_{0}^{\infty}d\alpha_i\frac{\mathcal{U}^{0}}{\mathcal{F}^2} \, , 
\end{align}
where both $\mathcal{U}$ and $\mathcal{F}$ are polynomials of degree $2$ and $3$, respectively, in the Schwinger parameters $\alpha_i$ see \cite{Panzer:2014caa} for more details\footnote{ The Schwinger parametrization of the integral $P_{1,34,25}$ can be obtained similarly.}. The integration in parametric space is possible for these integrals following an algorithm  introduced in \cite{Brown:2008um} and implemented in Maple in \cite{Panzer:2014caa}\footnote{We have also checked the results using the method of asymptotic expansions\cite{Smirnov:2002pj,Eden:2012rr,Goncalves:2016vir,Georgoudis:2017meq}. }. 
\begin{figure}[t]
\centering
\includegraphics[width=1.0\textwidth]{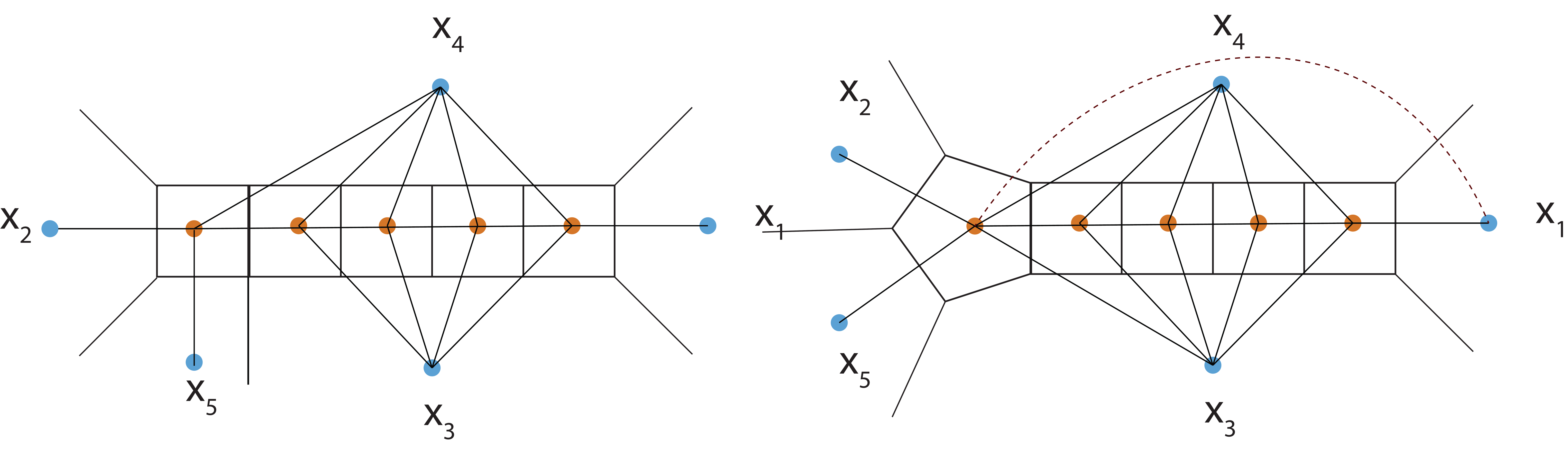}
\caption{Both position and momentum space representation of the ladder integral with $5$ points are represented on the left. On the right there is the analogue for the pentaladder integral. The dashed line represents a numerator.  }
\label{fig:figuraquenaoecitada} 
\end{figure}
We will now present a different approach to these integrals which can be also applied to their higher loop generalizations, see figure \ref{fig:figuraquenaoecitada}. Notice that both integrals satisfy a differential equation as the action of $\nabla^2_{2}$ or $\nabla_5^2$ localizes one of the integration  
\begin{align}
\nabla^2\frac{1}{x^2}= -4\pi^2\delta(x).
\end{align}
For example for the integral (\ref{eq:fivepointLadderintegral}) we have
\begin{align}
\frac{(x_{12}^2)^2x_{23}^2}{x_{13}^2}\nabla_2^2\left(\frac{L(z,\bar{z},w,\bar{w},h)}{x_{12}^2}\right)=\frac{x_{12}^2x_{15}^2x_{23}^2x_{34}^2}{x_{13}^2x_{24}^2}\int \frac{d^4x_7}{x_{16}^2x_{26}^2x_{36}^2x_{56}^2} \, .
\end{align}
 We are interested in the kinematics where all points are on the plane but this differential equation is derived assuming the points are at generic positions. Inserting the expansion of the integral $L$ around the perpendicular  distance $h$ out of the plane
\begin{align}
L(z,\bar{z},w,\bar{w},h)=\sum_{n=0}^{\infty}L^{(n)}(z,\bar{z},w,\bar{w})h^{n} \, , 
\end{align}
in the differential equation we see that the leading order in this expansion involves both $L^{(0)}$ and $L^{(1)}$. 

Acting with the Laplacian $\nabla_5^2$ on the integral $L$ gives another differential equation for $L(z,\bar{z},w,\bar{w},h)$
\begin{align}
\frac{(x_{15}^2)^3x_{23}^2x_{34}^2}{x_{13}^2}\nabla_5^2\left(\frac{L(z,\bar{z},w,\bar{w},h)}{x_{12}^2x_{15}^2x_{34}^2}\right)=\frac{(x_{15}^2)^2x_{23}^2x_{34}^2}{x_{13}^2x_{35}^2}\int\frac{d^4x_7}{x_{17}^2x_{27}^2x_{47}^2x_{57}^2} \, . 
\end{align}

It is possible to find a combination of each differential equations
\begin{align}
&\bigg[\frac{z\bar{z}(w\bar{w})^2(z-\bar{z})^2(1-z)(1-\bar{z})}{4}\nabla_2^2-\frac{w\bar{w}z\bar{z}(w-\bar{w})^2(1-z)(1-\bar{z})}{4}\nabla_5^2\bigg]L\\
&=\frac{(w\bar{w})^2(z-\bar{z})^2}{4}\int\frac{d^4x_7}{x_{17}^2x_{27}^2x_{47}^2x_{57}^2}-\frac{w\bar{w}z\bar{z}(1-z)(1-\bar{z})(w-\bar{w})^2}{4}\int\frac{d^4x_6}{x_{16}^2x_{26}^2x_{36}^2x_{56}^2} \, , 
\end{align}
such that the leading order in the expansion out of the plane only involves $L^{(0)}$
\begin{align}
&\bigg[(z-\bar{z})^2\partial_{z}\partial_{\bar{z}}-(w-\bar{w})^2\partial_{w}\partial_{\bar{w}}-\frac{z(z-\bar{z})}{\bar{z}}\partial{z}+\frac{\bar{z}(z-\bar{z})}{z}\partial_{\bar{z}}+\frac{w(w-\bar{w})}{\bar{w}}\partial_{w}\\
&-\frac{\bar{w}(w-\bar{w})}{w}\partial_{\bar{w}}\bigg]L^{0}=\frac{(z-\bar{z})^2w\bar{w}F^{(1)}\left(\frac{z(1-w)}{z-w},\frac{\bar{z}(1-\bar{w}}{\bar{z}-\bar{w}}\right)}{(w-z)(\bar{w}-\bar{z})}-\frac{(w-\bar{w})^2z\bar{z}F^{(1)}(\frac{z}{\bar{z}},\frac{\bar{z}}{\bar{w}})}{(1-w)(1-\bar{w})w\bar{w}}\nonumber
\end{align}
where 
\begin{align}
\int\frac{d^4x_6}{x_{16}^2x_{26}^2x_{36}^2x_{56}^2}=\frac{F^{(1)}\left(\frac{z(1-w)}{z-w},\frac{\bar{z}(1-\bar{w})}{\bar{z}-\bar{w}}\right)}{x_{13}^2x_{25}^2} \, .
\end{align}
The same strategy can be applied both to the pentaladder $P_{1,34,25}$ and to higher loop generalizations of these integrals.  It would be interesting to obtain the solution to these types of differential equations. 


\end{document}